# Uncertainty quantification and inverse modeling for subsurface flow in 3D heterogeneous formations using a theory-guided convolutional encoder-decoder network

**Rui Xu[1], Dongxiao Zhang[1, 2,\*], and Nanzhe Wang[3]**

[1] Intelligent Energy Laboratory, Peng Cheng Laboratory, Guangdong, P. R. China

[2] School of Environmental Science and Engineering, Southern University of Science and Technology, Guangdong, P. R. China

[3] College of Engineering, Peking University, Beijing, P. R. China

\* Corresponding author: Dongxiao Zhang (zhangdx@sustech.edu.cn)

**Abstract**

We build surrogate models for dynamic 3D subsurface single-phase flow problems with multiple vertical producing wells. The surrogate model provides efficient pressure estimation of the entire formation at any timestep given a stochastic permeability field, arbitrary well locations and penetration lengths, and a timestep matrix as inputs. The well production rate or bottom hole pressure can then be determined based on Peaceman's formula. The original surrogate modeling task is transformed into an image-to-image regression problem using a convolutional encoder-decoder neural network architecture. The residual of the governing flow equation in its discretized form is incorporated into the loss function to impose theoretical guidance on the model training process. As a result, the accuracy and generalization ability of the trained surrogate models are significantly improved compared to fully data-driven models. They are also shown to have flexible extrapolation ability to permeability fields with different statistics. The surrogate models are used to conduct uncertainty quantification considering a stochastic permeability field, as well as to infer unknown permeability information based on limited well production data and observation data of formation properties. Results are shown to be in good agreement with traditional numerical simulation tools, but computational efficiency is dramatically improved.

# 1 Introduction

The prediction of fluid transport in subsurface formations constitutes a non-trivial task (Song et al., 2016; Xu et al., 2020; Yin et al., 2020), partly due to incomplete knowledge of formation physical properties, such as porosity and permeability. Therefore, uncertainty quantification is usually desired prior to making production decisions (Zhang et al., 2000). Moreover, it is often necessary to infer unknown parameter fields based on available production/observation data (i.e., inverse modeling problems or history matching) (Chang et al., 2010) and to predict future production. The solution of both problems relies on accurate forward models that predict the dynamic pressure/saturation distribution based on given initial (IC) and boundary conditions (BC). Traditionally, such forward models are constructed by numerical approaches, such as finite difference, finite volume, or finite element methods (Aziz and Settari, 1979). These methods have successfully evolved over the past centuries and serve as reliable solvers for varieties of subsurface flow problems. However, they are not computationally efficient, especially when solving large-scale 3D problems on millions of grids, nor are they flexible in handling problems with complex physics. Surrogate models, on the other hand, are desirable alternatives since they improve the efficiency of forward prediction at a minimal cost of accuracy, by learning the correlation between inputs and outputs based on available data.

Recently, deep neural networks (DNNs) have been widely utilized to build surrogate models due to their strong capability to learn complex correlations between inputs and outputs (Goodfellow et al., 2016). Purely data-driven DNNs, however, possess limited generalization ability or they may make non-physical predictions, especially when the training data are of limited amount or of poor quality (LeCun et al., 2015). Theory-guided machine learning emerged under such conditions (Karpatne et al., 2017; Raissi et al., 2019). By incorporating physical laws, IC/BC constraints, engineering controls or expert knowledge into the training process of DNNs, the robustness and generalization ability of the trained models are dramatically improved. Varieties of theory-guided machine learning frameworks/models have been proposed, such as the theory-guided data science (TGDS) framework (Karpatne et al., 2017), the physics-informed neural network (PINN) (Raissi et al., 2019), and the theory-guided neural network (TgNN) (Wang et al., 2020) or its weak form variant (TgNN-wf) (Xu et al., 2021a, 2021b). Indeed, they have achieved promising results in the solution of partial differential equations (PDEs) (Raissi et al., 2019; Raissi and Karniadakis, 2018), inverse problems (Jo et al., 2019; Wang et al., 2020), uncertainty quantification (Tripathy and Bilionis, 2018; Wang et al., 2021a), and optimization tasks (Wang et al., 2022; Xu et al., 2021a).

Most of the works mentioned above constructed surrogate models based on fully-connected neural networks (FCNNs). It has been shown, however, that FCNNs have limited performance when building high-dimensional surrogate models (Mo et al., 2019a; Zhu and Zabaras, 2018). Furthermore, the large parameter space as a result of deep network structures required to deal with complex problems increases the computational burden. On the other hand, convolutional neural networks (CNNs) constitute promising alternatives. Image-to-image regression is more efficient to deal with high-dimensional problems. Moreover, the parameter-sharing scheme of convolutional kernels dramatically reduces the size of the parameter space. Recently, the data-driven convolutional encoder-decoder architecture has been frequently used to build surrogate models for subsurface flow and contaminant transport problems (Jin et al., 2020; Mo et al., 2020, 2019b; Zhu and Zabaras, 2018). However, few attempts have been made to incorporate theoretical guidance into the training process of CNNs. Zhu et al. (2019) proposed a physics-constrained encoder-decoder network that utilized the Sobel filter (Gao et al., 2010) to approximate the spatial gradients, but the temporal gradients could not be obtained. Therefore, only steady-state problems were studied in their work. Wang et al. (2021b) developed a theory-guided autoencoder (TgAE) framework that utilizes the discrete finite difference scheme to impose theoretical guidance, which can approximate both spatial and temporal gradients, but only 2D problems were probed without consideration of sources/sinks. Subsequently, they proposed a theory-guided CNN (TgCNN) framework based on TgAE, and studied 2D problems with consideration of producing wells (Wang et al., 2022, 2021c). In this work, we modify the TgCNN framework to solve 3D problems. The novelty of this work is three-fold: 1) we design a convolutional encoder-decoder network which makes use of 3D convolutions to capture local patterns not only in the horizontal direction, but also in the vertical direction; 2) the gravity effect is considered in the governing equation, and producing wells with varying spatial locations (in the horizontal direction) and arbitrary penetration lengths (in the vertical direction) can be handled effectively; and 3) we utilize parallel computing with multiple GPUs which can deal with input images (permeability fields) with high resolution efficiently. In this study, we use the modified TgCNN to build surrogate models for dynamic subsurface flow problems with local sinks (producing wells) using limited training data. Uncertainty quantification and inverse modeling tasks are investigated using the constructed surrogate models, which will be shown to have superior efficiency at a minimal cost of accuracy compared to conventional numerical simulation tools.

The remainder of this paper is organized as follows. In Section 2, we describe the forward and inverse problems formulation regarding 3D single-phase flow with multiple producing wells, which are commonly encountered in the early development stage of an oil formation. In Section 3, we briefly introduce the architecture of the convolutional encoder-decoder network for surrogate modeling and the method to incorporate theoretical guidance via discretizing the governing equations. We also demonstrate the inverse modeling technique (particle swarm optimization) used in this study. In Section 4, we demonstrate the construction of surrogate models based on TgCNN, and show their accuracy and extrapolation ability compared to data-driven CNNs. In Sections 5 and 6, we design several case studies using the surrogate models to solve uncertainty quantification and inverse modeling problems, and demonstrate the efficiency and accuracy of TgCNN compared to numerical simulation tools. In Section 7, we discuss limitations of the proposed method, suggest directions for future work, and conclude the study.

## 2 Problem description

### 2.1 Governing equations

We consider a petroleum formation saturated with oil at the early development stage drilled with multiple producing wells. The single-phase oil transport equation considering the gravity effect can be written as follows:

$$\nabla \cdot \left[ \rho \frac{k}{\mu} (\nabla p - \rho \mathbf{g}) \right] + \tilde{q} = \frac{\partial (\rho \phi)}{\partial t}, \tag{1}$$

where $\rho$ is the oil density; $k$ is the permeability; $\mu$ is the oil viscosity; $p$ is the oil pressure; $\mathbf{g}$ is the gravity factor; $\phi$ is the porosity; and $\tilde{q}$ is the mass flow rate of oil at the producing wells. By introducing the formation factor $B_o$, the oil compressibility factor $C_o$, and defining the oil potential $\Phi = p - \rho g z$, Eq. 1 can be rewritten in 3D:

$$\frac{\partial}{\partial x}\left(\frac{k_x}{\mu B_o}\frac{\partial \Phi}{\partial x}\right) + \frac{\partial}{\partial y}\left(\frac{k_y}{\mu B_o}\frac{\partial \Phi}{\partial y}\right) + \frac{\partial}{\partial z}\left(\frac{k_z}{\mu B_o}\frac{\partial \Phi}{\partial z}\right) + q_{sc} = \frac{\phi C_o}{B_o}\frac{\partial \Phi}{\partial t}, \qquad (2)$$

where $q_{sc}$ is the oil-producing volumetric flow rate at standard conditions.

## 2.2 Uncertainty quantification

When the parameters in Eq. 2 are random fields, a stochastic PDE is formulated, and the solution is no longer deterministic. Uncertainty quantification tasks arise, so that the statistics of the solution (e.g., mean and variance) can be estimated. Herein, we treat the permeability field $\mathbf{k}$ as a random field obeying a certain probability distribution function $P(\mathbf{k}) \in \Omega$. Let $\eta(\mathbf{x}, t, \mathbf{k(x)})$ denote the solution of the stochastic PDE. The mean and variance of $\eta$ can then be obtained as follows:

$$\mu(\mathbf{x}, t) = \int_\Omega \eta(\mathbf{x}, t, \mathbf{k(x)}) P(\mathbf{k})\, \mathrm{d}\mathbf{k} \qquad (3.1)$$

$$\sigma^2(\mathbf{x}, t) = \int_\Omega \left[\eta(\mathbf{x}, t, \mathbf{k(x)}) - \mu(\mathbf{x}, t)\right]^2 P(\mathbf{k})\, \mathrm{d}\mathbf{k} \qquad (3.2)$$

## 2.3 Inverse modeling

The prediction of dynamic pressure distribution given a realization of the permeability field at fixed IC/BC and well controls (WCs, e.g., producing at a certain flow rate or bottom hole pressure (BHP)) formulates the following forward problem:

$$f(k; IC, BC, WC) = p(\mathbf{x}, t), \qquad (4)$$

The corresponding inverse problem is defined as:

$$k = f^{-1}(p, IC, BC, WC) \qquad (5)$$

i.e., to resolve the unknown permeability field based on observation data of pressure, IC/BC, and WCs. In practice, however, the measurement of the pressure distribution of the entire formation is not feasible, and usually the observation data are limited to the vicinity of wells, such as well producing rates, BHPs, and local permeability data obtained by well logging or core analysis techniques. The inverse modeling task based on these limited data is therefore more challenging, and the solution might not be unique.

## 3 Methodology

### 3.1 Convolutional encoder-decoder network for surrogate modeling

We use a convolutional encoder-decoder network to approximate the complex correlation between the inputs and outputs of the forward model (Eq. 4), i.e., building surrogate models without solving the equation. The original surrogate modeling task in the continuous spatial and temporal domains is transformed into an image-to-image regression problem, by discretizing the problem on a mesh grid similar to that used in numerical approaches, such as the finite difference method. For the 3D problem in which we are interested, the regression function can be written as:

$$\eta: \mathbb{R}^{n_x \times H \times L \times W} \rightarrow \mathbb{R}^{n_y \times H \times L \times W}, \tag{6}$$

where $H$, $L$, and $W$ represent the number of grid points along the height, length, and width dimension, respectively; and $n_x$ and $n_y$ represent the number of input and output channels, respectively. Here, we have two input channels ($n_x = 2$), including the discretized permeability field ($k \in \mathbb{R}^{H \times L \times W}$) and the time matrix of the same size ($t \in \mathbb{R}^{H \times L \times W}$), with each element being the same time step $t_i$ at which the pressure is to be predicted. The output channel ($n_y = 1$) is the pressure field ($p \in \mathbb{R}^{H \times L \times W}$) at time step $t_i$.

The encoder-decoder network has been widely employed for image reconstruction and regression modeling tasks for image-type inputs and outputs (Goodfellow et al., 2016). Here, we use the autoencoder architecture (Wang et al., 2021b) shown in **Figure 1**, which is a special type of encoder-decoder network. The encoder component extracts high-level coarse features from input images by performing a sequence of convolutional operations. The fully connected layer transforms the output from the encoder component to a latent vector (i.e., the 'codes'). The decoder component then reconstructs the output based on the 'codes' by performing deconvolutional operations.

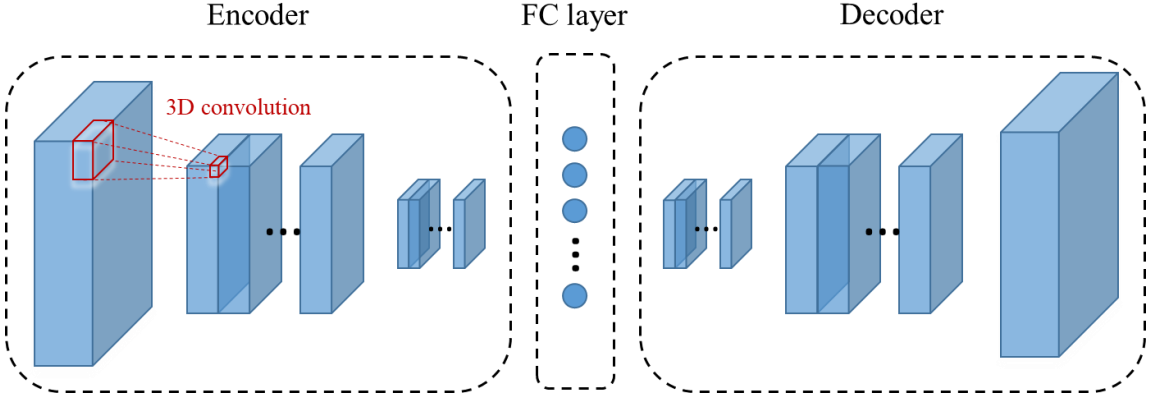

**Figure 1.** Schematic of the architecture of the autoencoder network used in this study.

The weights and biases of the convolutional filters and the fully connected layers are the major parameters (denoted as $\theta$) that need to be optimized during the training process, so that the following loss function is minimized:

$$Loss_{data}(\theta) = \sum_{i=1}^{N_k} \sum_{j=1}^{N_t} \|\eta(k_i, t_i) - \hat{\eta}(k_i, t_i; \theta)\|_2^2 \tag{7}$$

where $\eta$ and $\hat{\eta}$ are reference values and network predictions, respectively; and $N_k$ and $N_t$ are the total number of permeability field realizations and the number of timesteps used to calculate the pressure distributions, respectively, which constitutes the training dataset. The optimization task $\min Loss_{data}(\theta)$ can be conducted by different algorithms, such as stochastic gradient descent (SGD) or adaptive moment estimation (Adam) (LeCun et al., 2015).

## 3.2 Theory-guided training

The training process described in Section 3.1 is fully data-driven, and therefore highly depends on the amount and quality of the training data to achieve reasonable interpolation abilities. However, when there are limited or low-quality training data, the performance of data-driven models is not guaranteed. Here, we introduce the concept of theory-guided training in which known governing equations (PDEs) can be used to guide the training process (Raissi et al., 2019; Wang et al., 2020; Xu et al., 2021b). Specifically, the PDE residual approximated by the neural network can be added to the loss function and minimized during the training, so that the model prediction not only follows the patterns learned from the training data, but also obeys physical laws imposed by PDEs. As a result, the data dependence of the neural network training process is alleviated, which is especially beneficial in this case since 3D large-scale numerical simulations are time-consuming. The discretized Eq. 2 in the fully implicit scheme on the Cartesian grids of the input images can be written as:

$$
\left(\frac{k_x}{\mu B_o}\right)_{i+\frac{1}{2},j,k} \frac{\Phi_{i+1,j,k}^{n+1} - \Phi_{i,j,k}^{n+1}}{\Delta x^2} - \left(\frac{k_x}{\mu B_o}\right)_{i-\frac{1}{2},j,k} \frac{\Phi_{i,j,k}^{n+1} - \Phi_{i-1,j,k}^{n+1}}{\Delta x^2}
$$

$$
+ \left(\frac{k_y}{\mu B_o}\right)_{i,j+\frac{1}{2},k} \frac{\Phi_{i,j+1,k}^{n+1} - \Phi_{i,j,k}^{n+1}}{\Delta y^2} - \left(\frac{k_y}{\mu B_o}\right)_{i,j-\frac{1}{2},k} \frac{\Phi_{i,j,k}^{n+1} - \Phi_{i,j-1,k}^{n+1}}{\Delta y^2}
$$

$$
+ \left(\frac{k_z}{\mu B_o}\right)_{i,j,k+\frac{1}{2}} \frac{\Phi_{i,j,k+1}^{n+1} - \Phi_{i,j,k}^{n+1}}{\Delta z^2} - \left(\frac{k_z}{\mu B_o}\right)_{i,j,k-\frac{1}{2}} \frac{\Phi_{i,j,k}^{n+1} - \Phi_{i,j,k-1}^{n+1}}{\Delta z^2}
$$

$$
+ \frac{q_{sc}}{\Delta x \Delta y \Delta z} = \frac{\phi C_o}{B_o} \frac{\Phi_{i,j,k}^{n+1} - \Phi_{i,j,k}^{n}}{\Delta t} \tag{8}
$$

where the superscript $n$ indicates the timestep; subscripts $i$, $j$, and $k$ indicate the spatial coordinate; $\Delta x, \Delta y$, and $\Delta z$ are the grid intervals in $x, y$, and $z$ directions, respectively; and $\Delta t$ is the time interval. The inter-grid permeability is calculated as the harmonic mean of the two neighboring grids.

For vertical wells producing at a constant flow rate, the individual flow rate at each perforated grid is allocated based on its permeability, $q_i = q_{sc} k_i / \sum_{i=1}^{N_p} k_i$, where $k_i$ is the permeability of the $i$th perforated grid, and $N_p$ is the total number of perforated grids of that well along the vertical direction. For wells producing at a constant BHP, the flow rate of the $i$th perforated grid is calculated based on Peaceman's formula (Aziz and Settari, 1979):

$$
q_i = \frac{2\pi \Delta z \sqrt{k_x k_y}}{\mu \ln \frac{r_0}{r_w}} \left(p_{well}^i - BHP\right) \tag{9}
$$

where $p_{well}^i$ is the pressure of the $i$th perforated grid; $r_w$ is the wellbore radii; and $r_0$ is the radii of the effect drainage area, which is defined as follows:

$$
r_0 = 0.28 \frac{\sqrt{k_y \Delta x^2 + k_x \Delta y^2}}{\sqrt{k_x} + \sqrt{k_y}} \tag{10}
$$

The flow rate of the well is then the summation of all perforated grids. The PDE

residual ($R$) loss can be calculated from the neural network approximation of Eq. 8:

$$Loss_{PDE}(\theta) = \sum_{i=1}^{N_{k_v}} \sum_{j=1}^{N_{t_v}} \|R(k_i, t_i; \theta)\|_2^2 \tag{11.1}$$

$$R(k_i, t_i; \theta) = \left(\frac{k_x}{\mu B_o}\right)_{i+\frac{1}{2},j,k} \frac{\hat{\eta}(k_i, t_i; \theta)_{i+1,j,k} - \hat{\eta}(k_i, t_i; \theta)_{i,j,k}}{\Delta x^2}$$

$$- \left(\frac{k_x}{\mu B_o}\right)_{i-\frac{1}{2},j,k} \frac{\hat{\eta}(k_i, t_i; \theta)_{i,j,k} - \hat{\eta}(k_i, t_i; \theta)_{i-1,j,k}}{\Delta x^2}$$

$$+ \left(\frac{k_y}{\mu B_o}\right)_{i,j+\frac{1}{2},k} \frac{\hat{\eta}(k_i, t_i; \theta)_{i,j+1,k} - \hat{\eta}(k_i, t_i; \theta)_{i,j,k}}{\Delta y^2}$$

$$- \left(\frac{k_y}{\mu B_o}\right)_{i,j-\frac{1}{2},k} \frac{\hat{\eta}(k_i, t_i; \theta)_{i,j,k} - \hat{\eta}(k_i, t_i; \theta)_{i,j-1,k}}{\Delta y^2}$$

$$+ \left(\frac{k_z}{\mu B_o}\right)_{i,j,k+\frac{1}{2}} \frac{\hat{\eta}(k_i, t_i; \theta)_{i,j,k+1} - \hat{\eta}(k_i, t_i; \theta)_{i,j,k}}{\Delta z^2}$$

$$- \left(\frac{k_z}{\mu B_o}\right)_{i,j,k-\frac{1}{2}} \frac{\hat{\eta}(k_i, t_i; \theta)_{i,j,k} - \hat{\eta}(k_i, t_i; \theta)_{i,j,k-1}}{\Delta z^2}$$

$$+ \frac{q_{sc}}{\Delta x \Delta y \Delta z} = \frac{\phi C_o}{B_o} \frac{\hat{\eta}(k_i, t_i; \theta)_{i,j,k} - \hat{\eta}(k_i, t_i - \Delta t; \theta)_{i,j,k}}{\Delta t} \tag{11.2}$$

where $N_{k_v}$ and $N_{t_v}$ are the number of realizations of stochastic permeability fields and the number of time steps, respectively, used to enforce PDE constraints. It is worth noting that this set of realizations does not have labelled training data, but is only intended to calculate the PDE residual. In other words, the model learns the correlation between inputs and outputs in a semi-supervised manner, so that it can be applied to unseen permeability fields with similar statistics. Dirichlet and Neumann boundary conditions are incorporated in a soft manner, and the corresponding losses are defined as:

$$Loss_{BC}(\theta) = \sum_{i=1}^{N_{k_v}} \sum_{j=1}^{N_{t_v}} \left\| \nabla \hat{\eta}(k_i, t_i; \theta)|_{\Gamma_N} \cdot \vec{\mathbf{n}} - \mathbf{g} \right\|_2^2 + \left\| \hat{\eta}(k_i, t_i; \theta)|_{\Gamma_D} - h \right\|_2^2 \tag{12}$$

where $\Gamma_N$ and $\Gamma_D$ indicate the boundary grids that satisfy Neumann and Dirichlet BCs, respectively; and $\mathbf{g}$ and $h$ are corresponding Neumann and Dirichelt BCs.

The initial condition is imposed in a hard manner, via transforming the original network output $\hat{\eta}_0$ in the following way to obtain the updated output $\hat{\eta}$:

$$\hat{\eta} = \Phi_0(1-t) - t\hat{\eta}_0 \tag{13}$$

so that at initial state ($t = 0$), the network output is guaranteed to be the specified initial potential $\Phi_0$. The total loss is then defined as the summation of data mismatch, PDE residual, and BC constraints:

$$Loss = \lambda_1 Loss_{data}(\theta) + \lambda_2 Loss_{PDE}(\theta) + \lambda_3 Loss_{BC}(\theta) \tag{14}$$

where $\lambda_1$, $\lambda_2$, and $\lambda_3$ are the weights (Lagrange multipliers) of the corresponding losses to softly enforce the physical constraints.

## 3.3 Stochastic permeability fields generation

A series of stochastic permeability fields need to be generated to enforce theoretical guidance in the training process, so that a reliable surrogate model can be built that extrapolates to other permeability fields with similar statistics. Here, we consider Gaussian-type permeability fields, and use Karhunen–Loeve expansion (KLE) (Chang and Zhang, 2015; Zhang and Lu, 2004) to parameterize the random field. For a Gaussian random field $Z(\mathbf{x}, \tau) = \ln k(\mathbf{x}, \tau)$, where $\mathbf{x}$ is the spatial domain and $\tau$ is the probability space, it can be expressed as $Z(\mathbf{x}, \tau) = \bar{Z}(\mathbf{x}) + Z'(\mathbf{x}, \tau)$, where $\bar{Z}(\mathbf{x})$ is the average value of the random field and $Z'(\mathbf{x}, \tau)$ is the fluctuation. The spatial structure of the random field can be described by the two-point covariance function $C_Z(\mathbf{x}, \mathbf{x}') = \langle Z'(\mathbf{x}, \tau) Z'(\mathbf{x}', \tau) \rangle$. Using KLE, the random field can be decomposed as (Ghanem and Spanos, 2003):

$$Z(\mathbf{x}, \tau) = \bar{Z}(\mathbf{x}) + \sum_{i=1}^{\infty} \sqrt{\lambda_i} f_i(\mathbf{x}) \xi_i(\tau) \tag{15}$$

where $\lambda_i$ and $f_i$ are the $i$th eigenvalue and eigenfunction of the covariance, respectively; and $\xi_i$ is a Gaussian random variable with zero mean and unit variance. For separable exponential covariance function $C_Z(\mathbf{x}, \mathbf{x}') = \sigma_Z^2 \exp(-|x_1 - x_2|/\eta_x - |y_1 - y_2|/\eta_y - |z_1 - z_2|/\eta_z)$, where $\sigma_Z^2$ and $\eta$ are the variance and correlation length of the random field, respectively, the eigenvalues and eigenfunctions can be solved analytically or semi-analytically, the details of which can be found in Zhang and Lu (2004). Practically, we truncate Eq. 15 by the first several terms, and therefore a random field can be generated using a collection of $\xi$.

3.4 Particle swarm optimization for inverse problems

Particle swarm optimization (PSO) is a stochastic population-based optimization method proposed by Kennedy and Eberhart (1995). It simulates animals' social behavior, such as birds and fish, of food hunting in a cooperative way. Specifically, it solves an optimization problem by having a population of candidate particles moving around in the search-space to find the minimum fitness value (or the function to be minimized). Each particle's movement is influenced by its local best position, as well as the best positions of the entire species, so that the swarm is expected to move towards the best solutions. The major advantage of PSO compared to other gradient-based approaches, such as gradient descent or Newton methods, is that PSO does not require the problem to be differentiable (Bonyadi and Michalewicz, 2017). The position and velocity of the particles in PSO are iteratively updated based on the following formula:

$$v_i^{n+1} = \omega v_i^n + c_1 R_1^n (pbest_i^n - x_i^n) + c_2 R_2^n (gbest^n - x_i^n) \qquad (16.1)$$

$$x_i^{n+1} = x_i^n + v_i^{n+1} \qquad (16.2)$$

where $v_i^n$ and $x_i^n$ are the velocity and location of the $i$th particle at the $n$th iteration, respectively; $pbest_i^n$ is the local best known location of the $i$th particle; $gbest^n$ is the global best location of the entire species at the $n$th iteration; $c_1$ and $c_2$ are the learning rates; $R_1^n$ and $R_2^n$ are two sets of random numbers; and $\omega$ is the inertia weight, which is used to control the influence of the previous velocity value on the updated velocity (Shi and Eberhart, 1998).

**4 Surrogate models construction using TgCNN**

We consider a 3D formation saturated with oil. The geometrical and petrophysical properties of the formation and the oil within are shown in **Table 1.**

**Table 1.** Formation and oil properties.

| Property | Symbol | Value | Unit |
|---|---|---|---|
| formation top depth | $z_{top}$ | 3657.6 | m |
| formation length | $L_x$ | 365.76 | m |
| formation width | $L_y$ | 670.56 | m |
| formation height | $L_z$ | 51.82 | m |
| porosity | $\varphi$ | 0.2 | |
| oil density | $\rho_o$ | 849 | kg/m$^3$, STD |
| oil viscosity* | $v_o$ | 3 | mPa·s |
| oil compressibility* | $c_o$ | 0.0001 | bar$^{-1}$ |
| oil formation factor* | $B_o$ | 1.02 | |

**\* Note:** these properties are defined at a reference pressure of 413.69 bar at formation top depth.

We assume that the stochastic permeability field $K$ obeys a lognormal distribution with $\langle \ln K \rangle = 4$ mD, $\sigma^2_{\ln K} = 0.5$, and the correlation length is set to be $\eta_x = \eta_y = \eta_z = 152.4$ m. We truncate Eq. 15 with the first 13 terms, which reserves 60% of the energy. The random field can then be reconstructed using 13 independent random variables $\{\xi_i\}, i = 1, 2, \ldots, 13$. The 3D formation is discretized into 60, 220, and 10 grids in $x$, $y$, and $z$ directions, respectively. Four producing wells are placed near the four corners of the formation, and the wells penetrate and perforate the entire formation in the vertical direction. **Figure 2** shows three example realizations of permeability fields and the corresponding well locations. The initial pressure of the formation is gravitationally-stabilized, and the reference pressure at the top of the formation is 413.69 bar. The boundaries of the formation are treated as no-flow BCs. Two types of well control operations are considered, including constant flow rate (50 m$^3$/D, STD, Case 1) and constant BHP (350 bar, Case 2). We consider the calculation of the pressure distribution of the formation in 20 timesteps, with each timestep representing either 3 d (for Case 1) or 1 d (for Case 2). The reference solution of the problem is obtained using UNCONG software (Li et al., 2015), while many other numerical simulation tools can also be used.

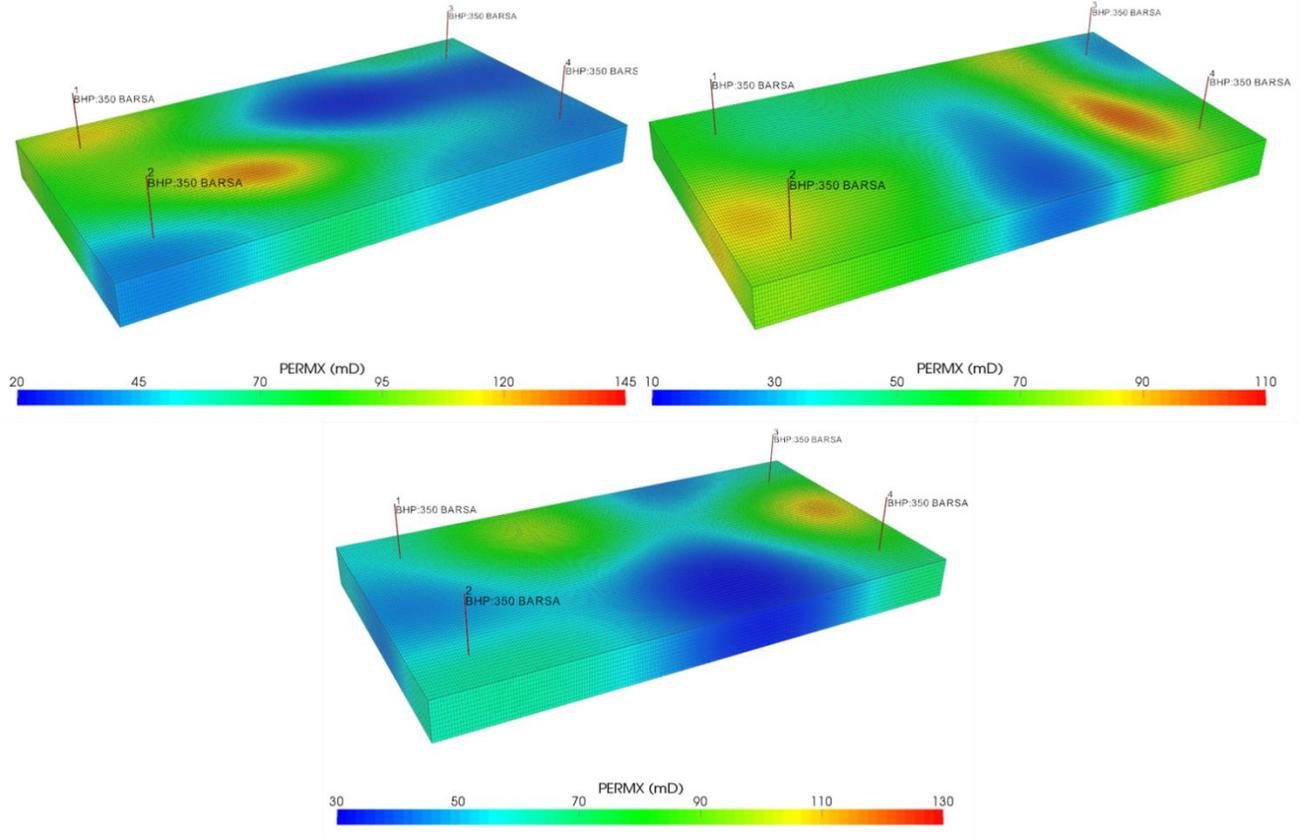

**Figure 2.** Three example realizations of the stochastic permeability field $K$ with $\langle ln\,K \rangle = 4$ mD, $\sigma_{ln\,K}^2 = 0.5$, and $\eta_x = \eta_y = \eta_z = 152.4$ m (visualizations obtained using UNCONG). The formation is discretized into 60, 220, and 10 grids in $x$, $y$, and $z$ directions, respectively. The locations of the four producing wells (Case 2 well controls are shown, and Case 1 has the same well locations) are shown, as well. The wells penetrate the entire formation in the vertical direction, and the spatial coordinates of the four wells in ($x$, $y$) grid number are as follows: (11, 21), (51, 21), (11, 201), and (51, 201).

The detailed convolutional encoder-decoder network architecture designed for the surrogate modeling task is shown in **Figure 3**. The input images consist of two channels, including the permeability field and time matrix. In the encoder component, 3D convolutional filters with different sizes and strides are applied, followed by the Swish activation function to reduce the input dimension and extract local features. A FC layer follows the encoder component with a single hidden layer and no activation function to encode the information into the latent variables. The decoder component performs a series of deconvolutional operations, followed by the ELU activation function to produce the output pressure images.

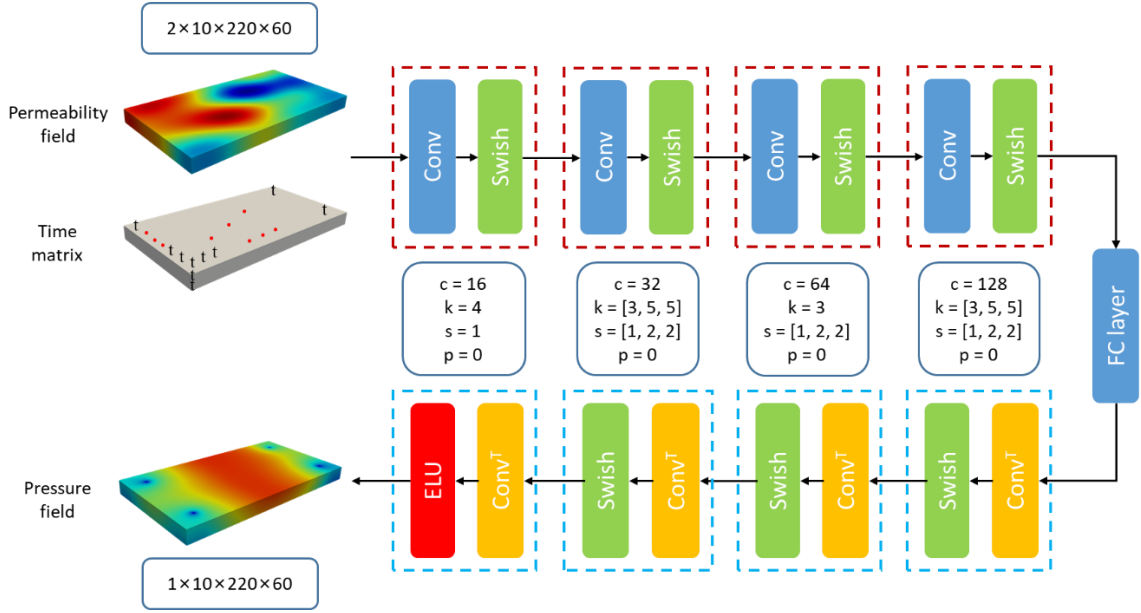

**Figure 3.** The encoder-decoder network architecture used for surrogate modeling. The parameters c, k, s, and p represent number of channels, convolutional kernel size, number of strides, and paddings, respectively.

## 4.1 Surrogate modeling for Case 1 (constant flow rate)

In this case, we consider the four wells, all producing at a constant flow rate of 50 m³/D, STD. The training parameters for Case 1 are listed in **Table 2**. The trained surrogate model is tested on 50 stochastic permeability fields generated using a different random seed from the training set. The accuracy of the surrogate model prediction compared to reference values is evaluated using two criteria, i.e., the relative L₂ error and the coefficient of determination (R² score), which are defined as follows:

$$L_2 = \frac{\|\eta^{NN} - \eta\|_2}{\|\eta\|_2}, \tag{17}$$

$$R^2 = 1 - \frac{\sum_{i=1}^{N_R} (\eta_i^{NN} - \eta_i)^2}{\sum_{i=1}^{N_R} (\eta_i - \bar{\eta})^2}, \tag{18}$$

where $\| \cdot \|_2$ represents the standard Euclidean norm; $N_R$ is the total number of evaluation points for $R^2$; and $\bar{\eta}$ is the average value of $\eta$, and $\eta$ can be either the pressure or well flow rate/BHP. The accuracy of the surrogate model is evaluated on each of the permeability fields in the training and testing sets, and the results are shown in the boxplots in **Figure 4**. For comparison, we also show the results using the fully data-driven CNN with the same network architecture and training parameters. As shown in **Figure 4**, both TgCNN and CNN have comparable accuracy and good extrapolation ability when evaluated on the entire pressure fields. CNN predictions of well BHP, however, are not accurate even on the training set, and the results are much worse than TgCNN on the testing set in terms of the mean and variance of the relative $L_2$ error or $R^2$ score. Therefore, incorporating theoretical guidance into the training process assists to capture local fluctuations (pressure disturbance caused by sources/sinks) more effectively. As an example, **Figure 5** shows the reference and TgCNN predicted pressure distribution at three timesteps for a permeability field drawn from the testing set (top left field in **Figure 2**). TgCNN predictions are close to the reference values with negligible absolute error. The largest error occurs at the corners of the formation. **Figure 6** shows the well BHP variation along the production timeline, and good agreement with the reference values is observed.

**Table 2.** Training parameters for Case 1.

| Training parameters | Value | Note |
|---|---|---|
| n_ln$k$_train | 5 | number of stochastic permeability fields used for pressure calculation as the training dataset |
| nt_train | 20 | number of time steps used to calculate the pressure distribution for the training dataset |
| n_ln$k$_virtual | 200 | number of stochastic permeability fields generated to enforce physical constraints during the training |
| n_batch | 100 | number of batches used to split the training data |
| lr | 0.001 | learning rate |
| $\lambda_1$ | 10 | weight for data loss |
| $\lambda_2$ | 0.3 | weight for PDE loss |
| $\lambda_3$ | 0.3 | weight for BC loss |
| loss_tol | 0.0004 | the tolerance value to terminate the training when the loss function drops below this value over the last 100 iterations |
| lr_decay_rate | 0.9 | decay rate of the learning rate |
| n_decay | 60 | the learning rate decays every n_decay epochs |
| epoch | 199 | final number of epochs |
| t_train | 38 min | training time (4 NVIDIA V100 GPUs trained in parallel) |

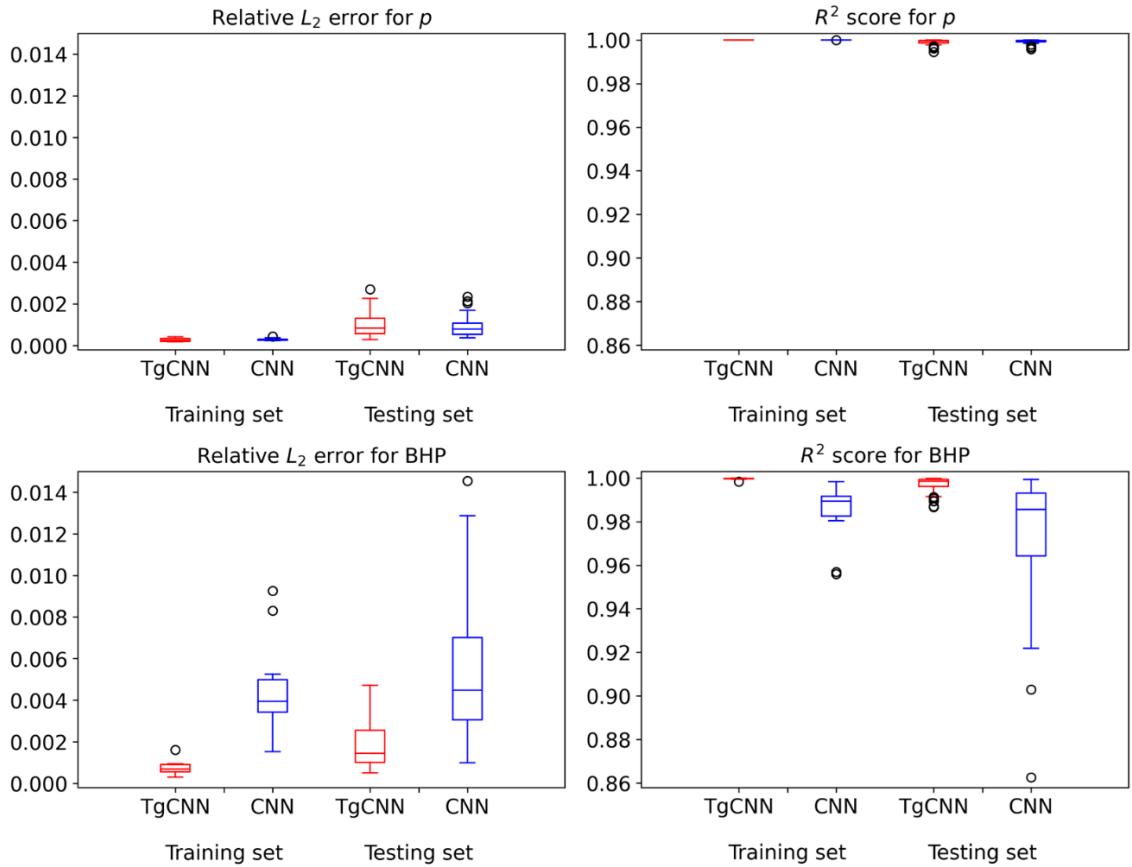

**Figure 4.** Relative $L_2$ error and $R^2$ score of the trained surrogate model (Case 1) evaluated on each of the permeability fields in the training set (five realizations) and the testing set (50 realizations) using TgCNN or CNN for the entire pressure field (first row) and the BHP of the four wells (second row).

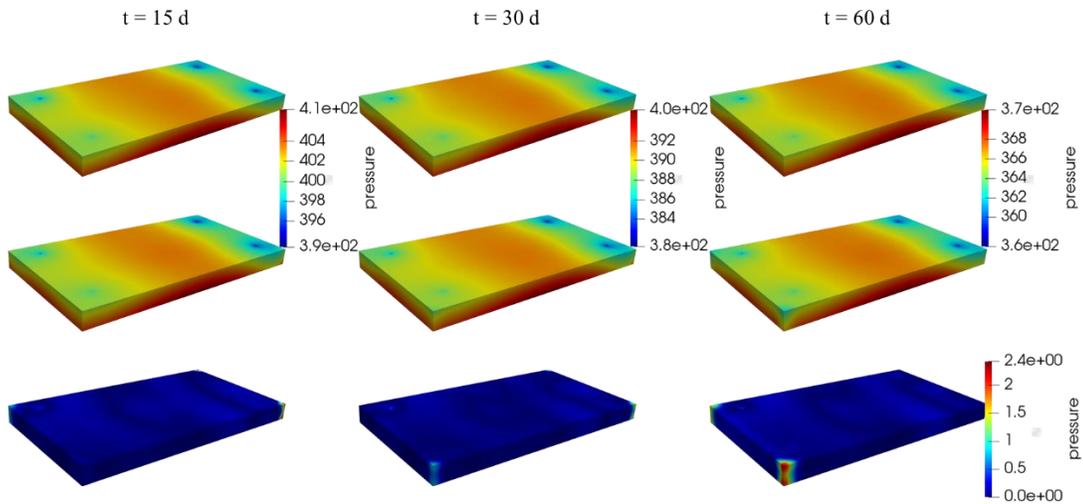

**Figure 5.** Case 1 pressure distribution visualization of a permeability field from the testing set (top left field in **Figure 2**) at three timesteps (15, 30, and 60 d) obtained by UNCONG software (reference, first row) and the TgCNN surrogate model (second row). The last row shows the absolute value of the absolute error between reference and predictions.

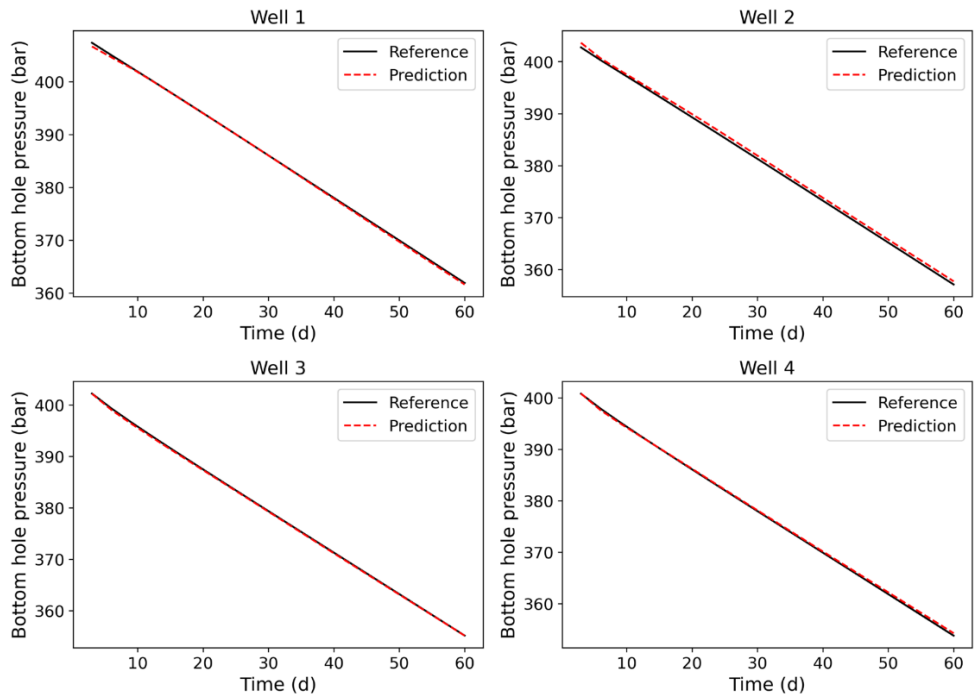

**Figure 6.** Case 1 well BHP variation along the production timeline for the four wells in a permeability field from the testing set (top left field in **Figure 2**) obtained by the TgCNN surrogate model compared to the reference values.

4.2 Surrogate modeling for Case 2 (constant BHP)

In this case, we consider the four wells, all producing at a constant BHP of 350 bar. The training parameters for Case 2 are listed in **Table 3**. The trained surrogate model is tested on 50 stochastic permeability fields generated using a different random seed from the training set. The accuracy of the surrogate model is evaluated on each of the permeability fields in the training and testing sets, and the results are presented in the boxplots in **Figure 7.** For comparison, results using the fully data-driven CNN are shown, as well. Different from Case 1, although both CNN and TgCNN have high accuracy on the training set for the entire pressure fields, the extrapolation ability of CNN to the testing set is poor with larger variance in accuracy. On the other hand, TgCNN has high accuracy on both the training and test sets (with $R^2$ score close to 1). For well flow rate prediction, the advantage of using TgCNN is more obvious. CNN has low accuracy on both the training and testing set (with a mean $R^2$ score of approximately 0.6, and several extreme cases with $R^2$ score close to zero), while TgCNN significantly outperforms CNN with $R^2$ score close to 1. This indicates that data-driven CNN surrogate models only ensure global accuracy at the cost of local accuracy, and possess limited extrapolation ability to inputs not seen previously. Incorporating theoretical guidance improves the models' generalization ability and assists the models to better capture local fluctuations/discontinuities. As an example, **Figure 8** shows the reference and TgCNN predicted pressure distribution at three timesteps for a permeability field drawn from the testing set (top left field in **Figure 2**). TgCNN predictions are close to the reference values with negligible absolute error. Similar to Case 1, the largest error occurs at the corners of the formation. **Figure 9** shows the well flow rate variation along the production timeline, and good agreement with the reference values is observed.

**Table 3.** Training parameters for Case 2.

| Training parameters | Value | Note |
|---|---|---|
| n_ln$k$_train | 30 | number of stochastic permeability fields used for pressure calculation as the training dataset |
| nt_train | 20 | number of time steps used to calculate the pressure distribution for the training dataset |
| n_ln$k$_virtual | 200 | number of virtual permeability fields generated to enforce physical constraints during the training |
| n_batch | 100 | number of batches used to split the training data |
| lr | 0.0005 | learning rate |
| $\lambda_1$ | 3 | weight for data loss |
| $\lambda_2$ | 0.03 | weight for PDE loss |
| $\lambda_3$ | 0.03 | weight for BC loss |
| loss_tol | 0.0008 | the tolerance value to terminate the training when the loss function drops below this value over the last 100 iterations |
| lr_decay_rate | 0.9 | decay rate of the learning rate |
| n_decay | 60 | the learning rate decays every n_decay epochs |
| epoch | 215 | final number of epochs |
| t_train | 45 min | training time (4 NVIDIA V100 GPUs trained in parallel) |

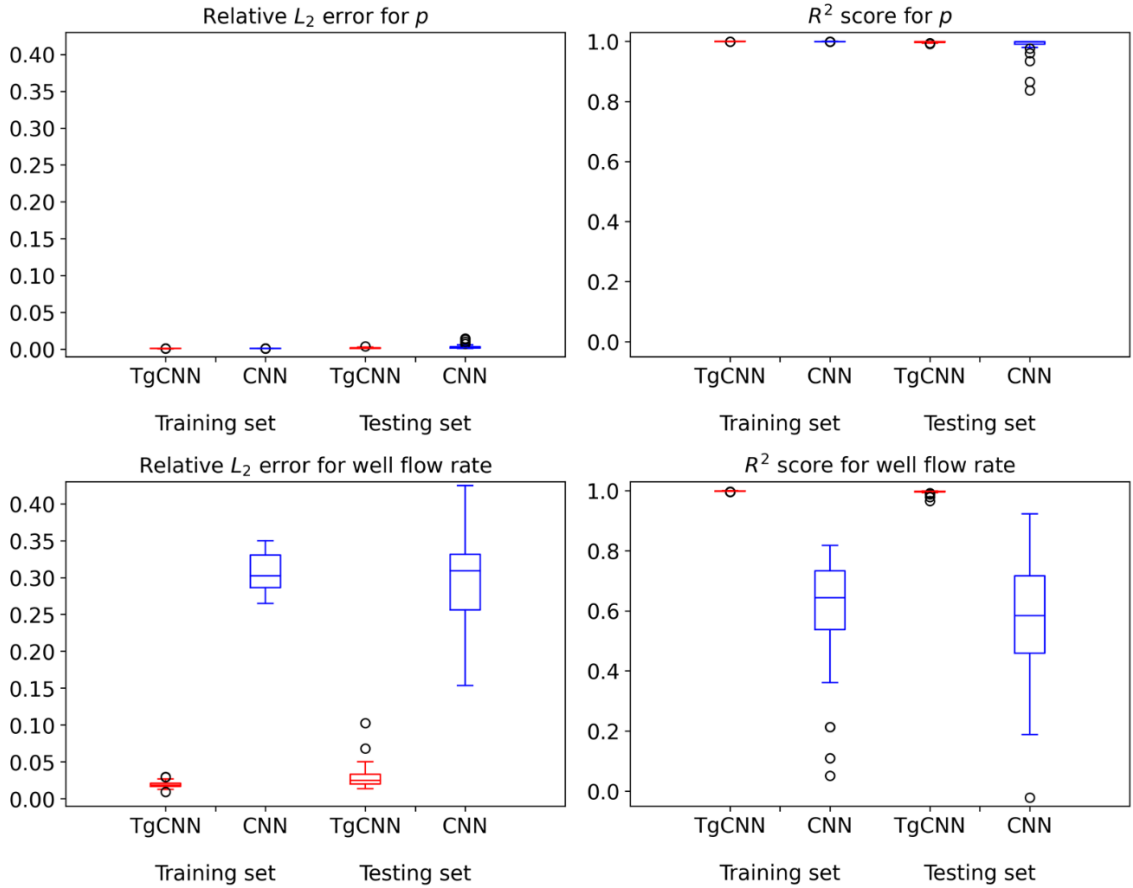

**Figure 7.** Relative $L_2$ error and $R^2$ score of the trained surrogate model (Case 2) evaluated on each of the permeability fields in the training set (30 realizations) and the testing set (50 realizations) using TgCNN or CNN for the entire pressure field (first row) and the flow rate of the four wells (second row).

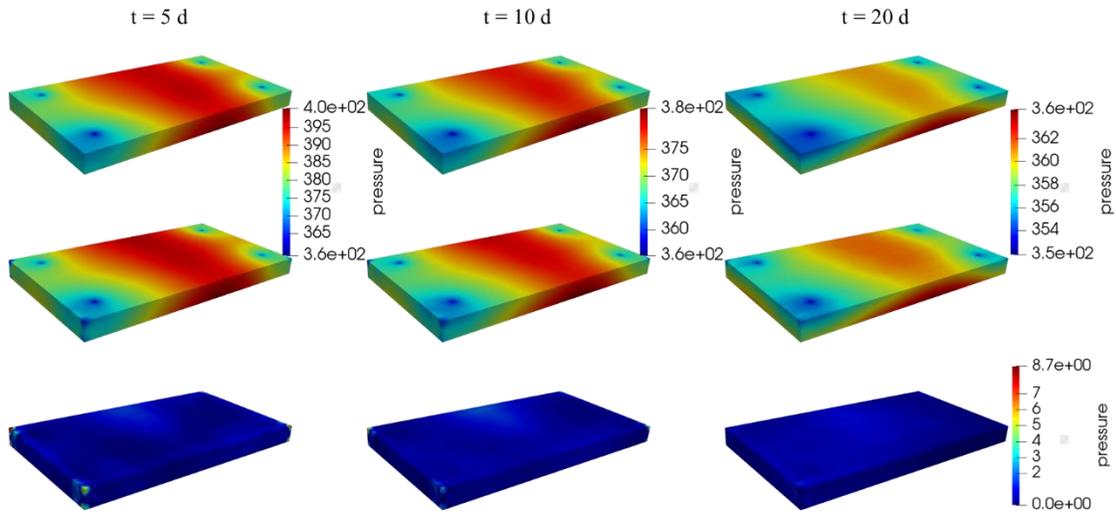

**Figure 8.** Case 2 pressure distribution visualization of a permeability field from the testing set (top left field in **Figure 2**) at three timesteps (5, 10, and 20 d) obtained by UNCONG software (reference, first row) and the TgCNN surrogate model (second row). The last row shows the absolute value of the absolute error between reference and predictions.

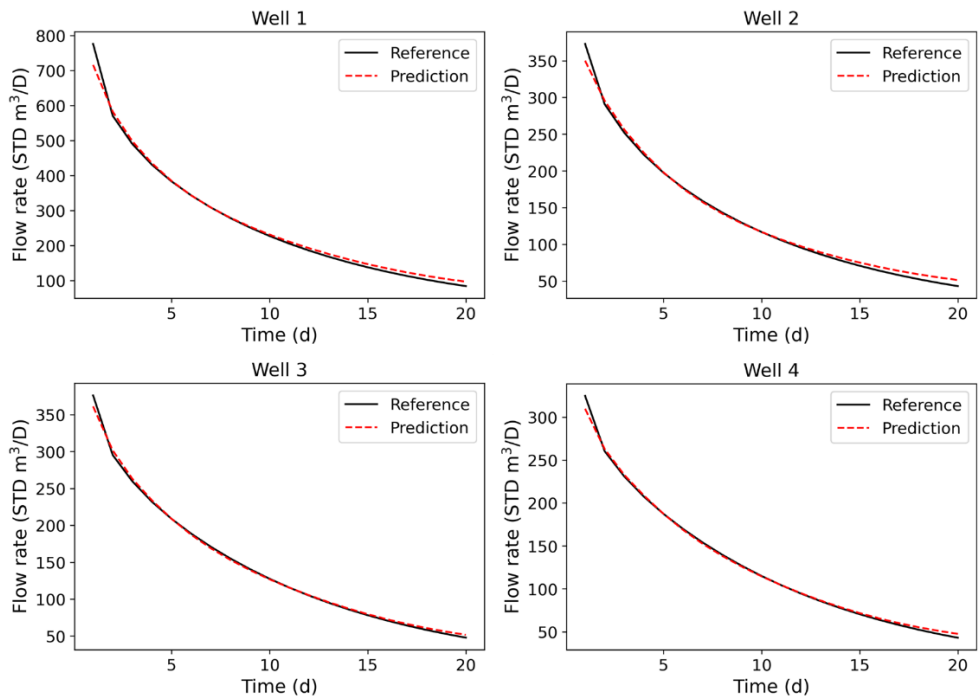

**Figure 9.** Case 2 well flow rate variation along the production timeline for the four wells in a permeability field from the testing set (top left field in **Figure 2**) obtained by the TgCNN surrogate model compared to the reference values.

### 4.3 Generalization ability of the surrogate models

Here, we test the generalization ability of the trained surrogate model to permeability fields with different statistics. For the sake of brevity, we only consider producing wells with constant BHP. We first keep the mean and correlation length of the stochastic permeability field to be constant (as used in Section 4.2, $\langle \ln K \rangle = 4$ mD, $\eta = 152.4$ m) and vary the variance of the permeability. Five cases are considered with $\sigma_{\ln K}^2 = 0.1, 0.3, 0.5, 0.7$, and $0.9$. For each case, 50 stochastic permeability fields are generated, and the pressure distribution over 20 timesteps is predicted using the same TgCNN surrogate model in Section 4.2 which was trained based on the case with $\sigma_{\ln K}^2 = 0.5$. The prediction accuracy of the pressure and well flow rates is quantified, as shown in **Figure 10**, in comparison to UNCONG simulation results. It can be seen that the generalization ability of the surrogate model to permeability fields with smaller variances ($\sigma_{\ln K}^2 < 0.5$) is much better than larger ones ($\sigma_{\ln K}^2 > 0.5$), which is as expected since permeability fields with larger variances introduce stronger heterogeneity and the surrogate model may struggle to handle out-of-distribution inputs (permeability values). In general, the accuracy of the surrogate model increases with the decrease in $\sigma_{\ln K}^2$, and the surrogate model can be trusted to provide reliable predictions for permeability fields with $\sigma_{\ln K}^2 < 0.7$ (with $R^2 > 0.98$).

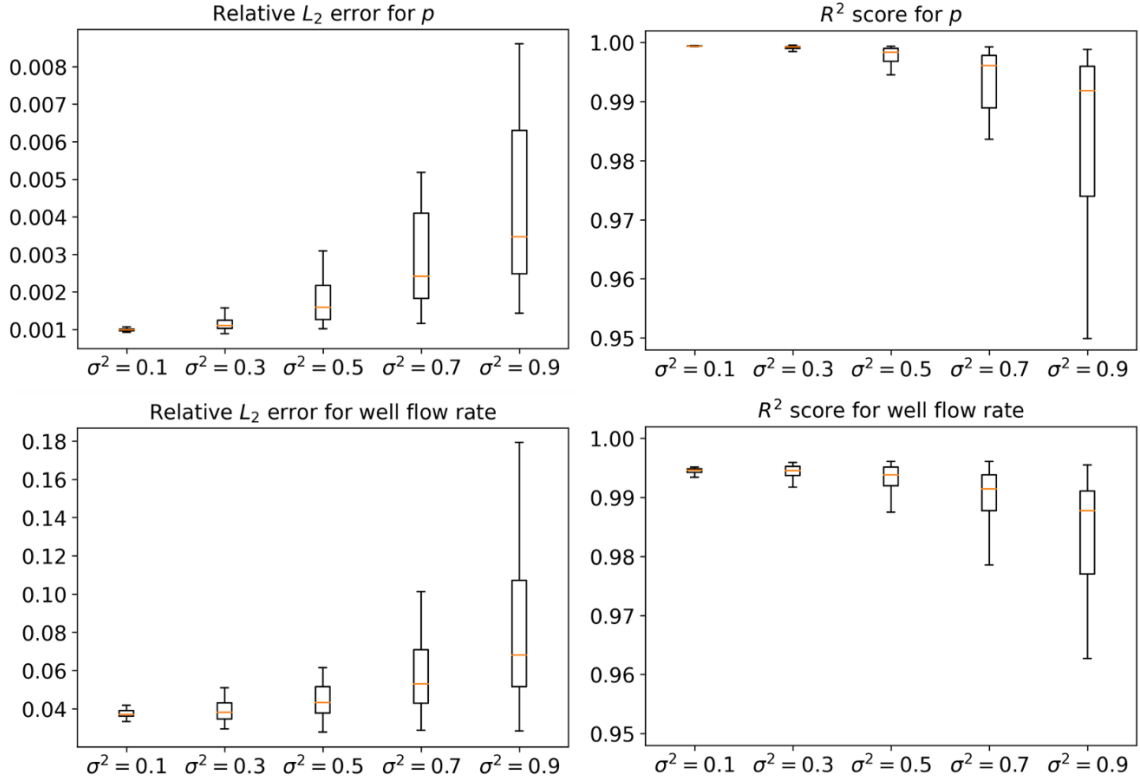

**Figure 10.** Relative $L_2$ error and $R^2$ score of the trained surrogate model evaluated on each one of the 50 permeability fields with $\langle \ln K \rangle = 4$ mD, $\eta = 152.4$ m, and varying variances $\sigma_{\ln K}^2 = 0.1, 0.3, 0.5, 0.7,$ and $0.9$. Results for the entire pressure field are shown in the first row, and results for the flow rates of the four wells are shown in the second row.

We then keep the mean and variance of the stochastic permeability field to be constant ($\langle \ln K \rangle = 4$ mD, $\sigma_{\ln K}^2 = 0.5$) and vary the correlation length. Five cases are considered with $\eta = 100, 130, 152, 170,$ and $200$ m. For each case, 50 stochastic permeability fields are generated, and the pressure distribution over 20 timesteps is predicted using the same TgCNN surrogate model (trained based on the case with $\eta = 152$ m). The prediction accuracy of the pressure and well flow rates is quantified, as shown in **Figure 11**, in comparison to UNCONG simulation results. The generalization ability of the surrogate model to permeability fields with larger correlation length ($\eta > 152$ m) is much better than smaller ones ($\eta < 152$ m), which is as expected since permeability fields with smaller correlation length introduce more complicated local patterns that are not learned by the convolutional filters. In general, the accuracy of the surrogate model increases with the correlation length, and the surrogate model can be trusted to provide reliable predictions for permeability fields with $\eta > 130$ m (with $R^2 > 0.93$).

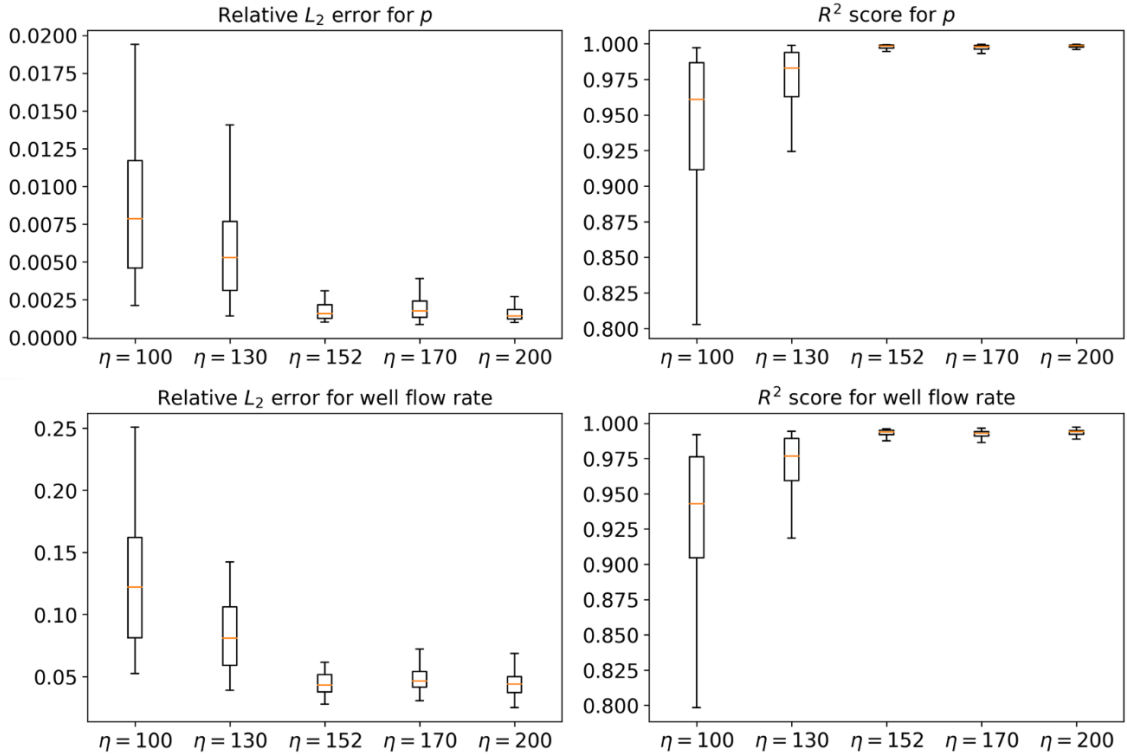

**Figure 11.** Relative $L_2$ error and $R^2$ score of the trained surrogate model evaluated on each one of the 50 permeability fields with $\langle ln\,K \rangle = 4$ mD, $\sigma^2_{ln\,K} = 0.5$, and varying correlation lengths $\eta = 100$, 130, 152, 170, and 200 m. Results for the entire pressure field are shown in the first row, and results for the flow rates of the four wells are shown in the second row.

### 4.4 Surrogate modeling for varying well locations and penetration lengths

Here, we consider a more complex situation, in which it is desired to estimate the pressure profile and well production rates (assuming constant BHP) given not only a stochastic permeability field, but also arbitrary well locations and well penetration lengths (the total number of wells remain the same). In this way, a more general surrogate model can be constructed to provide efficient uncertainty quantification and inverse modeling tasks for different well patterns, without the need to retrain the model. In this case, an addition input channel is needed, which provides well locations and penetration lengths information. Here, we construct a 3D well image which is a binary image of the same size as the permeability image. The image grids containing the wells (of arbitrary location and penetration length) take the value of one, while other grids take the value of zero. The network structure is the same as previous cases with only an additional input channel. The training parameters are listed in **Table 4**.

**Table 4.** Training parameters for the more complex surrogate model.

| Training parameters | Value | Note |
|---|---|---|
| n_train | 100 | number of stochastic permeability fields used for pressure calculation as the training dataset |
| n_well_train | 100 | number of well images used for pressure calculation as the training dataset |
| nt_train | 20 | number of time steps used to calculate the pressure distribution for the training dataset |
| n_ln$k$_virtual | 500 | number of virtual permeability fields generated to enforce physical constraints during the training |
| n_well_virtual | 500 | number of virtual well images generated to enforce physical constraints during the training |
| n_batch | 200 | number of batches used to split the training data |
| lr | 0.0005 | learning rate |
| $\lambda_1$ | 3 | weight for data loss |
| $\lambda_2$ | 0.3 | weight for PDE loss |
| $\lambda_3$ | 0.3 | weight for BC loss |
| loss_tol | 0.0008 | the tolerance value to terminate the training when the loss function drops below this value over the last 100 iterations |
| lr_decay_rate | 0.9 | decay rate of the learning rate |
| n_decay | 30 | the learning rate decays every n_decay epochs |
| epoch | 400 | final number of epochs |
| t_train | 1.5 h | training time (4 NVIDIA V100 GPUs trained in parallel) |

The trained surrogate model is tested on 50 stochastic permeability fields, with random well locations and penetration lengths generated using a different random seed from the training set. The overall $R^2$ for the pressure and well production rates are 0.994 and 0.996, respectively, indicating good extrapolation ability of the surrogate model. As an example, we demonstrate in **Figure 12** the comparison of pressure and well production rates estimation between the surrogate model and UNCONG simulation for three different testing cases, with detailed well location and penetration length information presented in **Table 5**. For all three cases, TgCNN-predicted pressure and well production rates closely align with the reference values, indicating good generalizability of the surrogate model not only to permeability fields with similar statistics, but also to arbitrary combinations of well locations and penetration lengths.

**Table 5.** Well locations and penetration lengths for the three testing cases shown in **Figure 12**.

| Case | Well coordinates (in grid numbers) | | | | Penetration lengths (in grid numbers) | | | |
|---|---|---|---|---|---|---|---|---|
| | Well 1 | Well 2 | Well 3 | Well 4 | Well 1 | Well 2 | Well 3 | Well 4 |
| 1 | (17, 177) | (19, 44) | (21, 56) | (16, 144) | 3 | 1 | 4 | 3 |
| 2 | (19, 137) | (45, 123) | (52, 133) | (47, 127) | 7 | 8 | 6 | 10 |
| 3 | (36, 213) | (35, 17) | (52, 131) | (7, 177) | 3 | 5 | 2 | 5 |

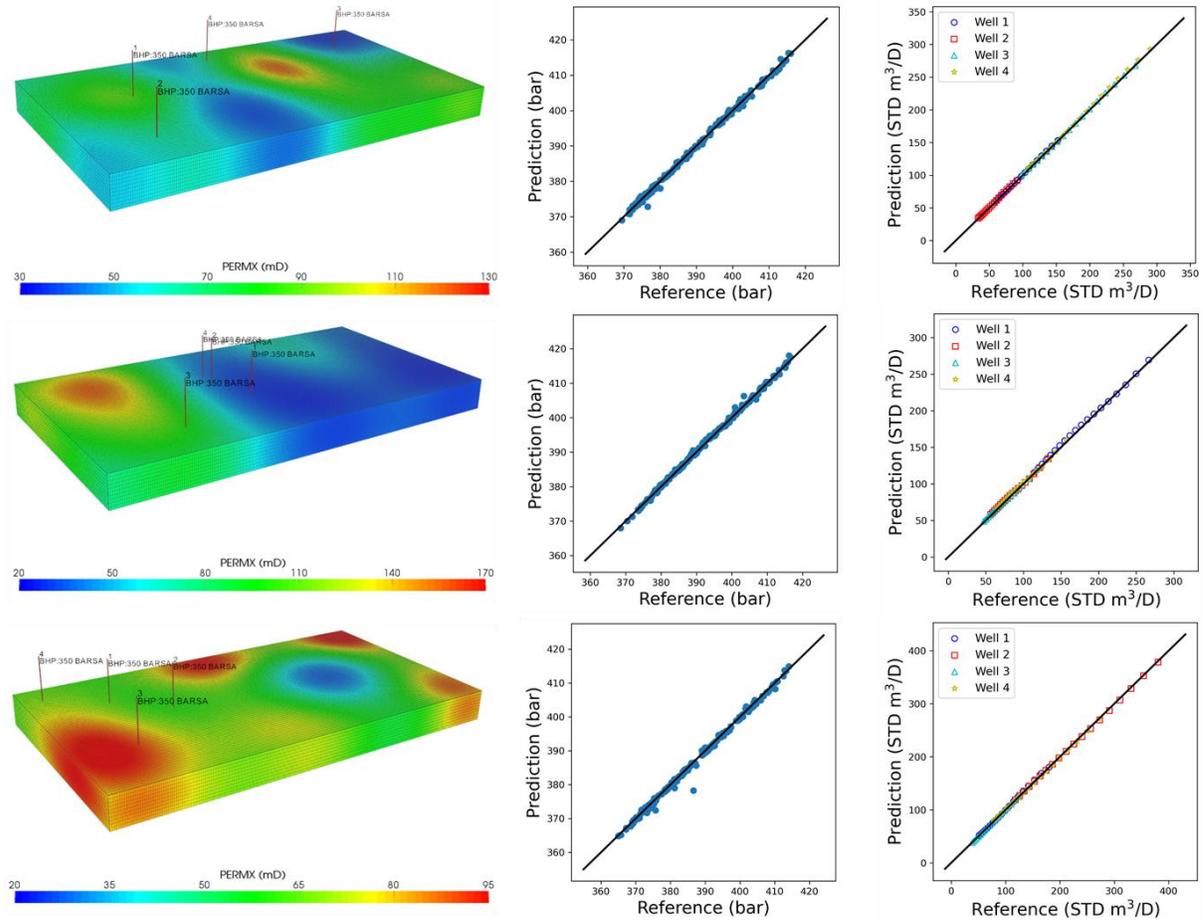

**Figure 12.** Comparison of pressure (second column, randomly extracted from 200 spatial-temporal locations) and four wells' production rates (third column) between the reference (UNCONG simulation) and TgCNN surrogate model prediction for three testing cases with stochastic permeability fields and well locations (the corresponding penetration lengths can be found in **Table 5**) shown in the first column.

## 5 Uncertainty quantification

Using the constructed surrogate model introduced in Section 4, we perform uncertainty quantification of the pressure of the entire field, producing well BHPs (for Case 1) and well flow rates (for Case 2). We assume that the statistics of the permeability field are known beforehand with $\langle \ln K \rangle = 4$ mD, $\sigma_{\ln K}^2 = 0.5$, and the correlation length is $\eta_x = \eta_y = \eta_z = 152.4$ m. The well locations and penetration lengths are the same as in Section 4.1. The statistical results are analyzed using 2,000 randomly-generated permeability fields. The same procedures are also performed using the Monte Carlo method with UNCONG software, the results of which are taken as the reference values.

For Case 1, the prediction of pressure distribution of 2,000 permeability fields in 20 timesteps takes approximately 9 min for the TgCNN surrogate model, while it takes more than 11 h for UNCONG to finish the same calculation. The efficiency of the surrogate model, even with consideration of training time (38 min), is more than 10 times faster than traditional numerical simulation tools, such as UNCONG. Uncertainty quantification results (mean and variance) of pressure in the entire field at three timesteps are shown in **Figure 13**. Good agreement with UNCONG results is observed with relatively large errors near the wells and around the corners. **Figure 14** shows the well BHP uncertainty analysis of the four producing wells. At each timestep, the mean and variance of the estimated BHP by the surrogate model and UNCONG software closely align with each other. The trained surrogate model exhibits significantly improved efficiency with satisfactory accuracy when solving uncertainty quantification problems.

For Case 2, the prediction of pressure distribution of 2,000 permeability fields in 20 timesteps takes approximately 9.5 min for the TgCNN surrogate model, while it takes more than 13 h for UNCONG to finish the same calculation. Uncertainty quantification results (mean and variance) of pressure in the entire field at three timesteps is shown in **Figure 15**. In this case, the accuracy of the mean and variance estimation is more accurate than Case 1, with the largest error around the corners of the formation. **Figure 16** shows the well flow rate uncertainty analysis of the four producing wells, and the surrogate model also has slightly better performance than Case 1, with higher accuracy in mean and variance predictions.

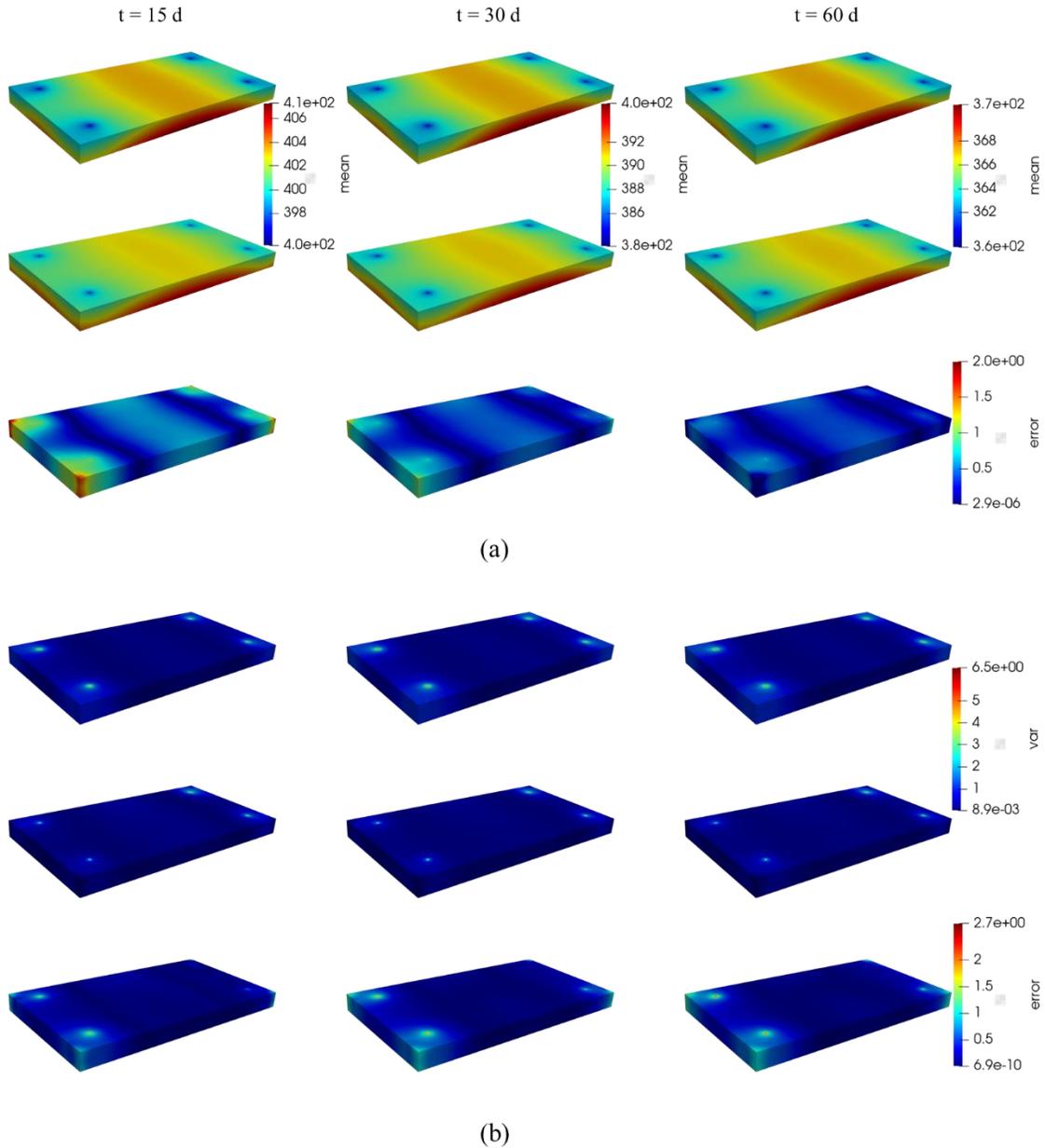

**Figure 13.** Case 1 uncertainty quantification of the (a) mean and (b) variance of the dynamic pressure of the entire formation considering a stochastic permeability field with $\langle ln\,K \rangle = 4$ mD, $\sigma_{ln\,K}^2 = 0.5$, and $\eta_x = \eta_y = \eta_z = 152.4$ m, at three timesteps (15, 30, and 60 d) obtained by UNCONG software (reference, first row) and the TgCNN surrogate model (second row). The last row in each subfigure shows the absolute value of the absolute error between reference and predictions.

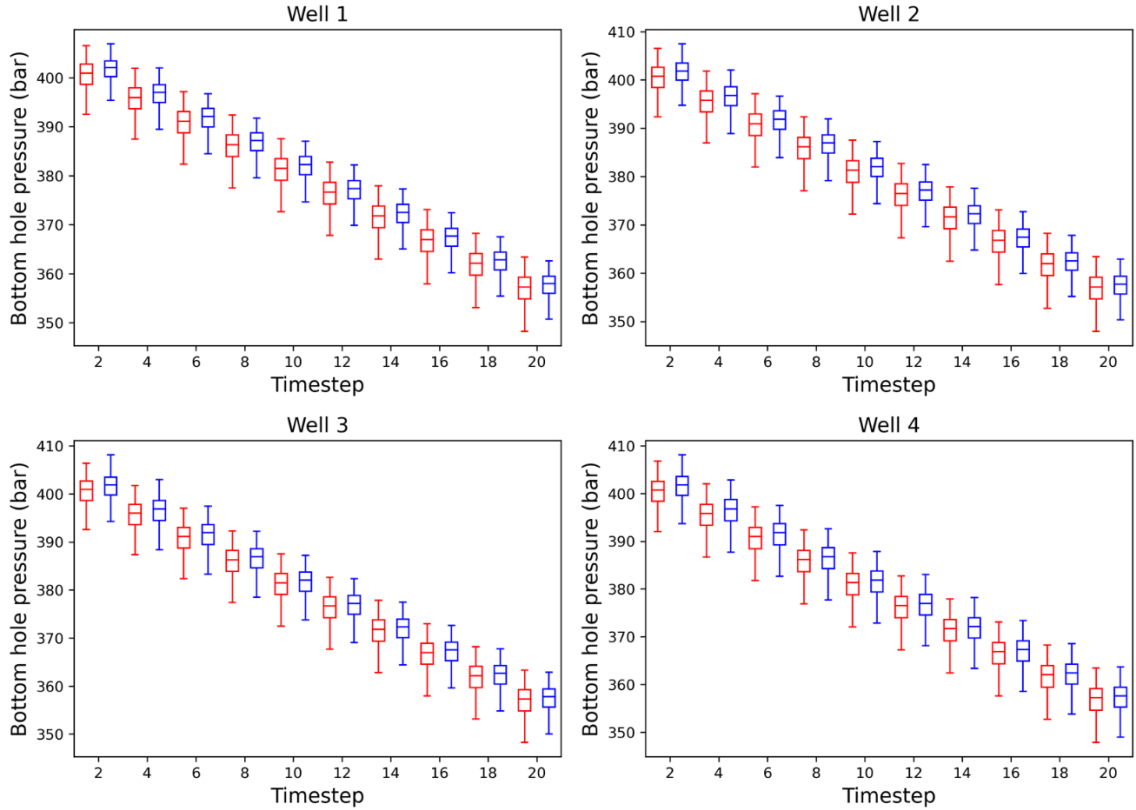

**Figure 14.** Uncertainty quantification of well BHP for Case 1 at different timesteps using the TgCNN surrogate model (blue) and UNCONG simulation (red). The results are calculated based on 2,000 randomly-generated permeability fields with $\langle ln\,K \rangle = 4$ mD, $\sigma^2_{ln\,K} = 0.5$, and $\eta_x = \eta_y = \eta_z = 152.4$ m.

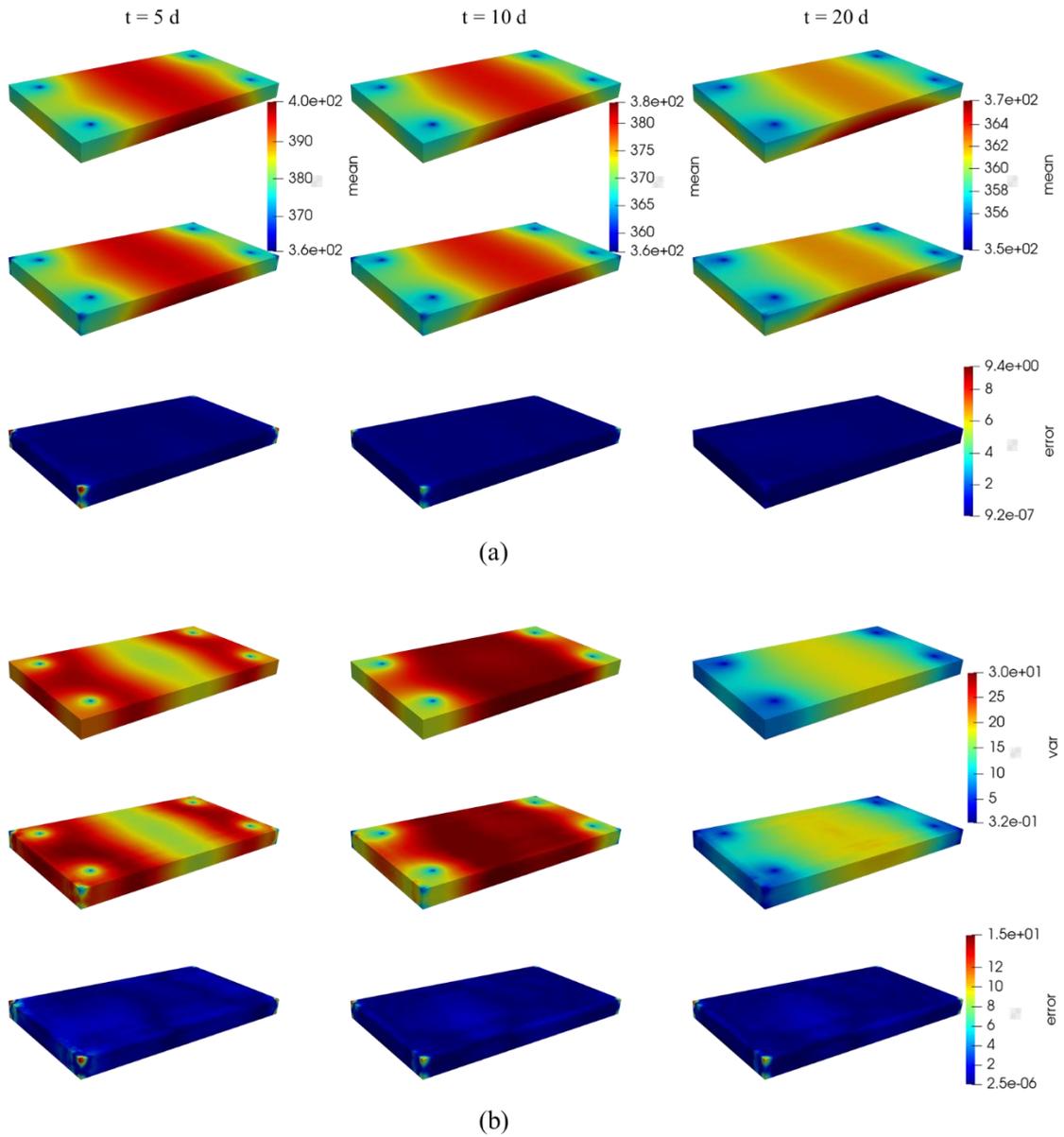

**Figure 15.** Case 2 uncertainty quantification of the (a) mean and (b) variance of the dynamic pressure of the entire formation considering a stochastic permeability field with $\langle ln\,K \rangle = 4$ mD, $\sigma_{ln\,K}^2 = 0.5$, and $\eta_x = \eta_y = \eta_z = 152.4$ m, at three timesteps (5, 10, and 20 d) obtained by UNCONG software (reference, first row) and the TgCNN surrogate model (second row). The last row in each subfigure shows the absolute value of the absolute error between reference and predictions.

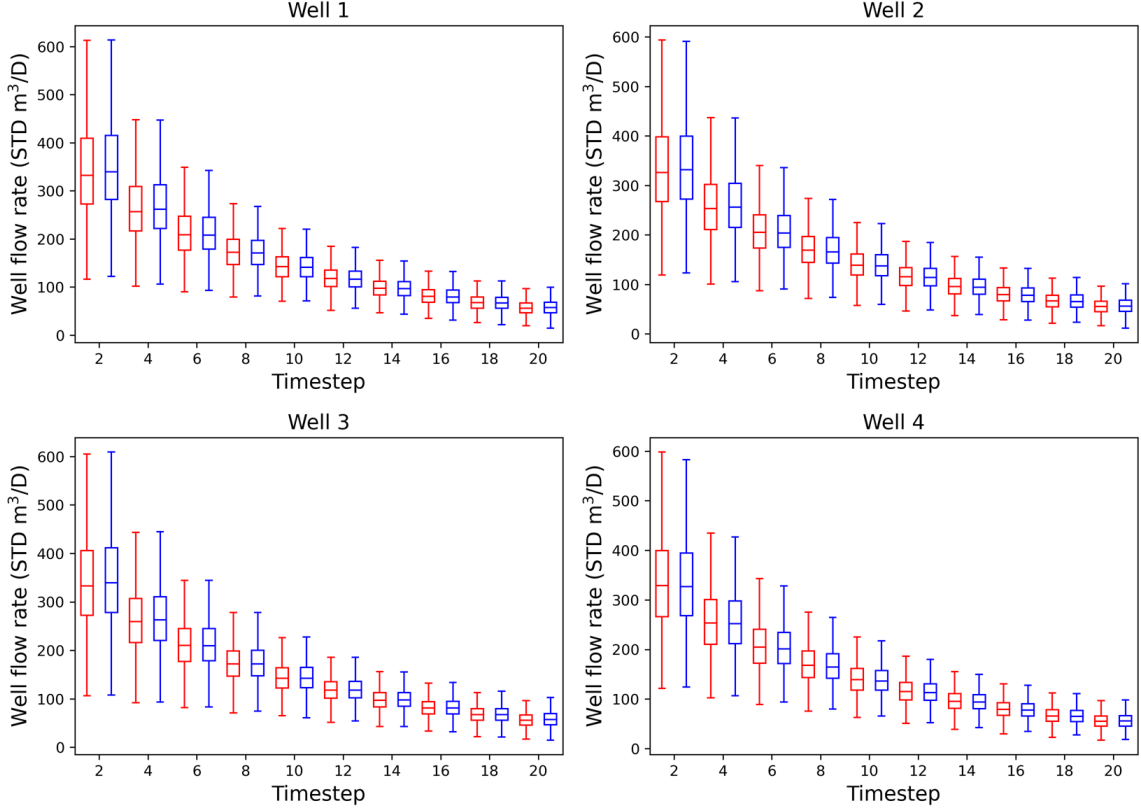

**Figure 16.** Uncertainty quantification of well flow rates for Case 2 at different timesteps using the TgCNN surrogate model (blue) and UNCONG simulation (red). The results are calculated based on 2,000 randomly-generated permeability fields with $\langle ln\,K \rangle = 4$ mD, $\sigma_{ln\,K}^2 = 0.5$, and $\eta_x = \eta_y = \eta_z = 152.4$ m.

## 6 Inverse modeling

We consider the inverse modeling problem for Case 2 to infer the permeability field based on observation data of well production rates, well BHP, known statistics of the permeability field, and local known permeability of the perforated grids along the well direction which may be obtained by logging tools or core analysis.

We start with a simple case in which the statistics of the permeability field are known to be $\langle \ln K \rangle = 4$ mD, $\sigma_{\ln K}^2 = 0.5$, and the correlation length is $\eta_x = \eta_y = \eta_z = 152.4$ m. The well production data are assumed to have been recorded in the first 10 d (10% noise is added for practical consideration, $\Delta t = 1$ d, total production time = 20 d), and the permeability of the grids penetrated by the wells is assumed to be known. The goal is to infer the permeability field based on available data and predict the well production in the next 10 d. The parameters used in the PSO algorithm are presented in **Table 6**. The optimization function (fitness value) is defined as follows:

$$FV = \frac{\lambda_1}{N_t N_{well}} \sum_{i=1}^{N_t} \sum_{j=1}^{N_{well}} \left( q_j^i - q_{ref}{}_j^i \right)^2 + \frac{\lambda_2}{N_{well} N_k} \sum_{i=1}^{N_{well}} \sum_{j=1}^{N_k} \left( k_j^i - k_{ref}{}_j^i \right)^2$$

$$+ \frac{\lambda_3}{N_t N_{well}} \sum_{i=1}^{N_t} \sum_{j=1}^{N_{well}} \left( BHP_j^i - BHP_{ref}{}_j^i \right)^2 \tag{19}$$

where $q$ is the well flow rate; $k$ is the permeability of the well grid; $N_k$ is the total number of grids for a well that has permeability measurement data, here $N_k$=10; the subscript '*ref*' indicates the observed/known value; and $\lambda$ is the weight of each term, subject to the numerical value of the term and the reliability of the observation. Here, we set $\lambda_1 = 1, \lambda_2 = 1000$, and $\lambda_3 = 10$ . The parameters to be optimized are the 13 random numbers ($\{\xi_i\}, i = 1,2,...,13$) used to generate the permeability field via KLE. The initial population of the swarm particles is generated randomly, and the optimization terminates in 50 steps, which takes approximately 12 min using a single GPU. The variation of the fitness value during the optimization process is shown in **Figure 17**. The fitness value converges rapidly in the first several iterations. The reference and inferred permeability fields are shown in **Figure 18**. Although noisy observation data in limited spatial and temporal domains are taken for the inverse modeling problem, the inverted permeability field is generally in good agreement with the reference. Particularly, the high permeability zones in the upper left corner and middle right are captured accurately. Using the optimized permeability field, we calculate the flow rates of the four wells in 20 d, and the results are shown in **Figure 19**. Good agreement with the reference data in the first 10 d is observed, although the observation data are quite noisy. Furthermore, the prediction of well production in the last 10 d is also close to the reference, indicating the reliability of the permeability field inversion, at least for well production estimation.

**Table 6.** Parameters used in the PSO algorithm for the case with known permeability statistics.

| Parameters | Value | Note |
|:---:|:---:|:---|
| $\omega$ | 0.9 | inertia weight |
| $c_1$ | 2 | learning rate for the personal best term |
| $c_2$ | 2 | learning rate for the global best term |
| maxgen | 30 | maximum number of iterations |
| sizepop | 20 | size of the population in the swarm |
| vmax | 1 | maximum particle velocity |
| vmin | -1 | minimum particle velocity |
| xmax | 4 | upper bound of particle location |
| xmin | -4 | lower bound of particle location |

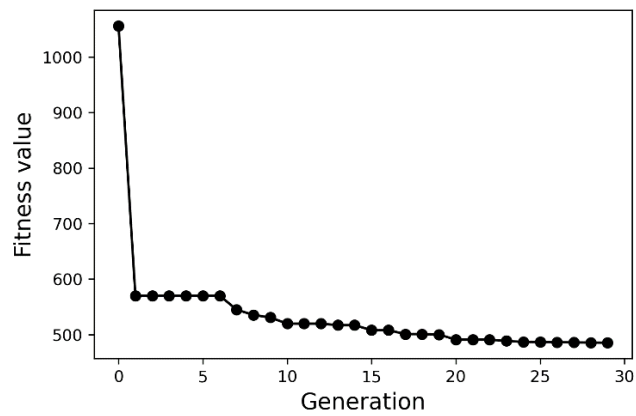

**Figure 17.** Variation of the fitness value of the global best particle during the PSO optimization process for the case with known permeability statistics.

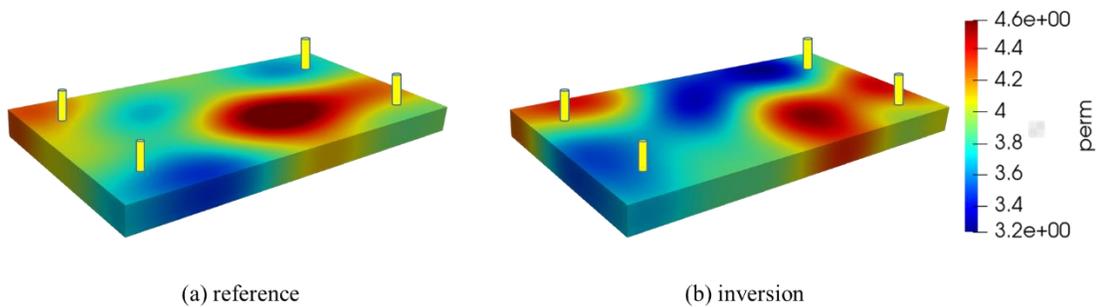

(a) reference            (b) inversion

**Figure 18.** Reference and inferred permeability fields for the case with known permeability statistics. The yellow cylinders indicate the location of the wells where known information of producing rates, pressure, and permeability is taken.

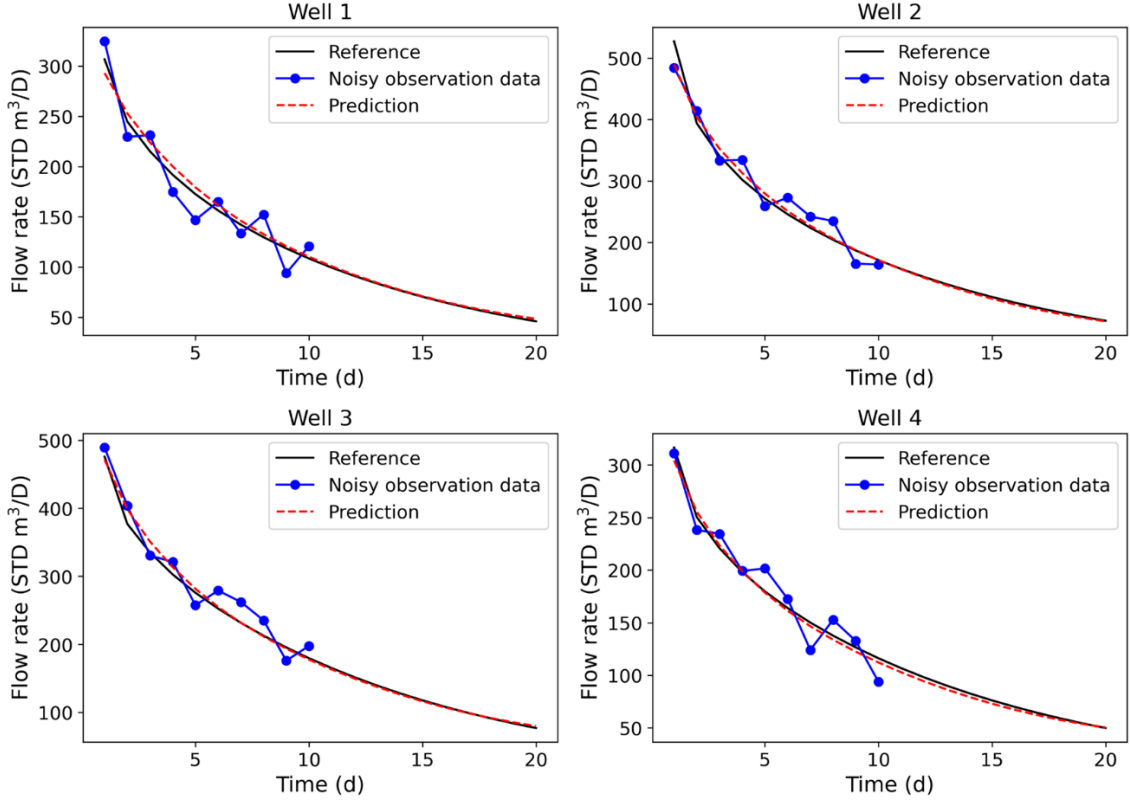

**Figure 19.** The flow rates of the four wells predicted using the inferred permeability field, in contrast to the noisy observation data and reference data (for the case with known permeability statistics).

We then consider a case in which only the mean value of the permeability field is known to be $\langle \ln K \rangle = 4$ mD, while the variance and correlation length are unknown, which shall also be inferred from the observation data. Here, we generate the observation data using a realization of the stochastic permeability field with $\sigma_{\ln K}^2 = 0.6$ and $\eta_x = \eta_y = \eta_z = 170$ m. Similarly, the flow rate data from the first 10 d are extracted and biased with 10% noise as the observation data. The parameters to be optimized include the random variable $\{\xi_i\}$, $\sigma_{\ln K}^2$, and $\eta$. Similar parameters to **Table 6** are used in the PSO algorithm in this case, except that we increase the size of the particle population and iteration steps to 30 and 50, respectively, to increase the probability of finding the global minimum, and we constrain the searching limits for $\sigma_{\ln K}^2$ and $\eta$ to be $[0, 1]$ and $[130, 190]$, respectively. The optimization process takes approximately 2 h, and the variation of the fitness value during the optimization process is shown in **Figure 20.** The reference and inferred permeability fields are shown in **Figure 21**. In this case, the inferred permeability field closely resembles the reference, except in the middle region where there are no observation data. The inferred values for $\sigma_{\ln K}^2$ and $\eta$ are 0.53 and 180, respectively, both of which are close to the reference values. Using the optimized permeability field, we calculate the flow rates of the four wells in 20 d, and the results are shown in **Figure 22**. Good agreement with the reference production data in the first 10 d is observed despite the noisy observation data, and the prediction of well production in the last 10 d is accurate, as well. Although part of the statistics of the permeability field is unknown, which dramatically increases the size of the searching space, the TgCNN-based surrogate model combined with the PSO algorithm provides satisfactory inversion results efficiently.

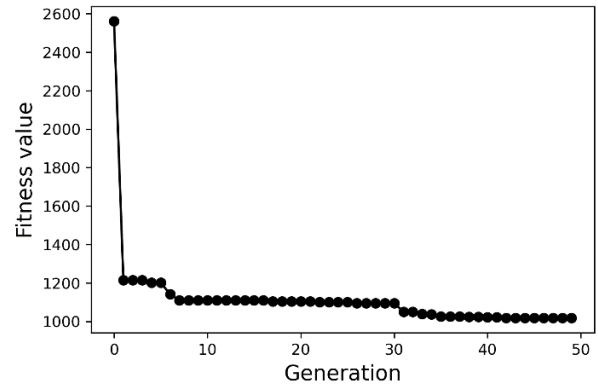

**Figure 20.** Variation of the fitness value of the global best particle during the PSO optimization process for the case with unknown permeability statistics.

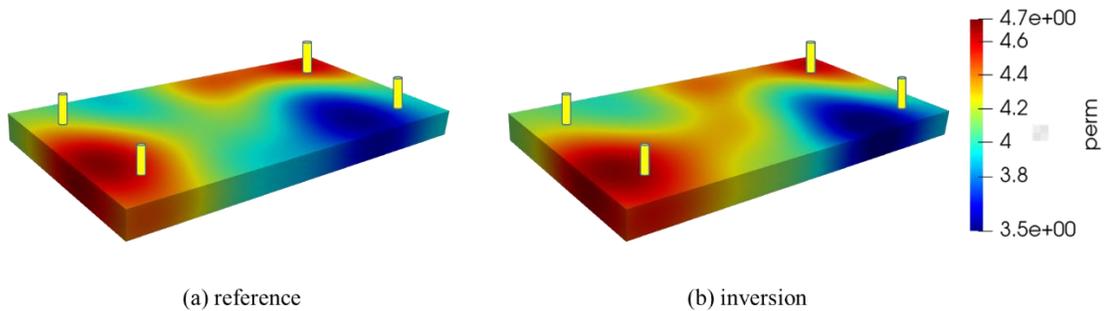

**Figure 21.** Reference and inferred permeability fields for the case with unknown permeability statistics. The yellow cylinders indicate the location of the wells where known information of producing rates, pressure, and permeability is taken.

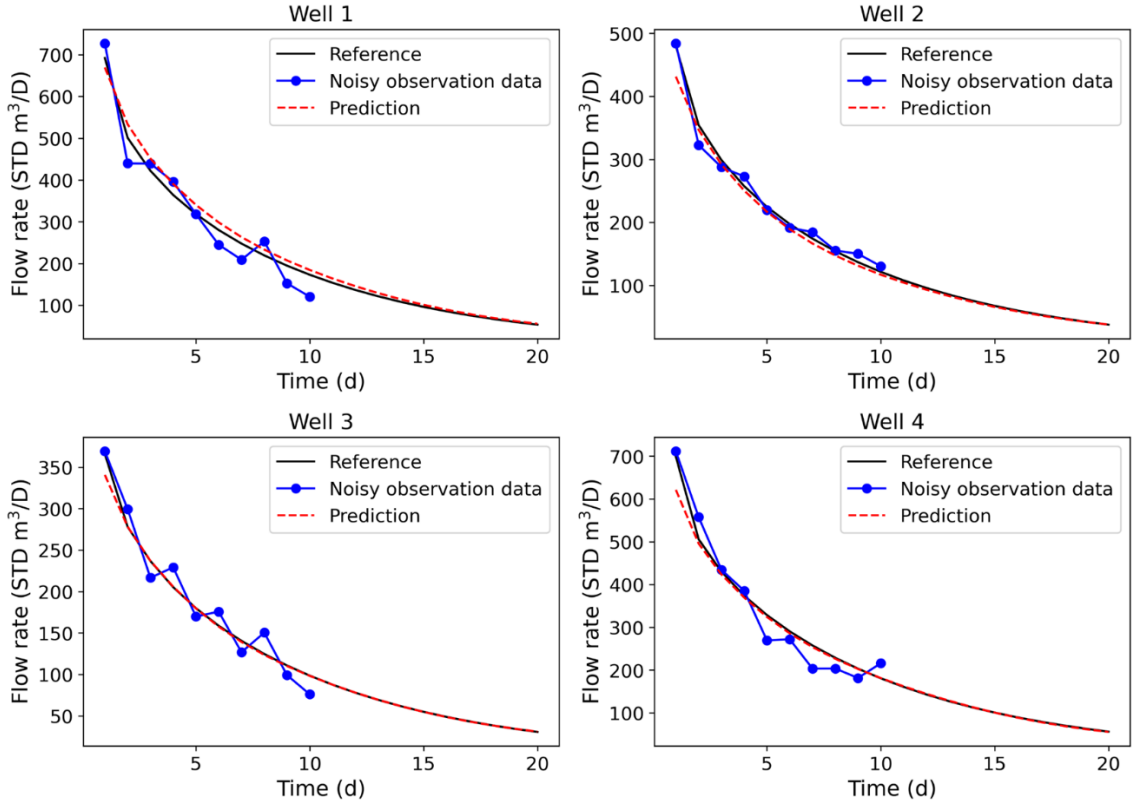

**Figure 22.** The flow rates of the four wells predicted using the inferred permeability field, in contrast to the noisy observation data and reference data (for the case with unknown permeability statistics).

## 7 Conclusions

We present the use of a theory-guided convolutional encoder-decoder neural network to build surrogate models for 3D dynamic subsurface fluid flow problems in a heterogeneous and stochastic permeability field, with consideration of multiple producing wells at arbitrary locations with arbitrary penetration lengths. Theoretical guidance in the form of discretized PDEs is constructed naturally using the input image grids, and different well production controls (constant pressure or flow rate) can be handled effectively. The surrogate models trained with theoretical guidance show improved accuracy and generalization ability compared to fully data-driven CNNs. Moreover, TgCNN-based surrogate models show good extrapolation ability to permeability fields with different statistics. The uncertainty quantification examples demonstrate the superior computational efficiency of the surrogate model (tens of times faster than numerical simulation tools, such as UNCONG) at a negligible cost of accuracy. The surrogate models combined with the PSO algorithm are also shown to constitute reliable tools for inverse modeling tasks. In general, TgCNN-based surrogate models are efficient to train, and accurate in prediction or extrapolation.

In this study, we consider simple Gaussian permeability fields with relatively small variance. In practice, however, the permeability fields may be more heterogeneous with larger variance or smaller correlation length, and high-permeability channels might be present. The resulting flow behavior is therefore more complicated to predict, and requires a more sophisticated neural network architecture. We defer treatment of more complicated (non-Gaussian) permeability fields to future work. Although only vertical wells are considered in this study, the treatment of horizontal wells is straightforward and will be explored in future research. Furthermore, here we consider single-phase flow with only producing wells at the early development stage of an oil formation. After the depletion of natural energy, water is usually injected to maintain pressure and displace oil, which results in two-phase or multi-phase flow problems. Surrogate models for such problems should predict not only pressures, but also saturations, of the formation. Neural networks with multiple outputs or multiple neural networks can be used for this purpose. We will address two-phase flow in future work.

**Acknowledgement**

We acknowledge the Peng Cheng Cloud Brain at the Peng Cheng Laboratory for providing high performance GPU computational resources.